\documentclass{aa}

\usepackage[varg]{txfonts}
\usepackage{natbib,graphicx}

\bibpunct{(}{)}{;}{a}{}{,} % to follow the A&A style

\begin{document}

\title{ Solar photosphere magnetization}

\author{V\'eronique Bommier}

\institute{LESIA, Observatoire de Paris, Universit\'e PSL, CNRS, Sorbonne Universit\'e, Universit\'e de Paris 
\newline 5, Place Jules Janssen, 92195 Meudon, France}

\date{Received ... / Accepted ...}

\abstract
% context
{A recent review shows that observations performed with different telescopes, spectral lines, and interpretation methods all agree about a vertical magnetic field gradient in solar active regions on the order of 3 G/km, when a horizontal magnetic field gradient  of only 0.3 G/km  is found. This represents an inexplicable discrepancy with respect to the $\mathrm{div}\vec{B}=0$ law.}
% aims
{The objective of this paper is to explain these observations through the law $\vec{B}=\mu_0\left(\vec{H}+\vec{M}\right)$ in magnetized media.}
% methods
{Magnetization is due to  plasma diamagnetism, which results from the spiral motion of free electrons or charges about the magnetic field. Their usual photospheric densities lead to very weak magnetization $\vec{M}$, four orders of magnitude lower than $\vec{H}$. It is then assumed that electrons escape from the solar interior, where their thermal velocity is much higher than the escape velocity, in spite of the effect of protons. They escape from lower layers in a quasi-static spreading, and accumulate in the photosphere. By evaluating the magnetic energy of an elementary atom embedded in the magnetized medium obeying the macroscopic law $\vec{B}=\mu_0\left(\vec{H}+\vec{M}\right)$, it is shown that the Zeeman Hamiltonian is due to the effect of $\vec{H}$. Thus, what is measured is $\vec{H}$.}
% results
{The decrease in   density with height is responsible for non-zero divergence of $\vec{M}$, which is compensated for by the divergence of $\vec{H}$, in order to ensure $\mathrm{div}\vec{B}=0$. The behavior of the observed quantities is recovered.}
% conclusions
{The problem of the divergence of the observed magnetic field in solar active regions finally reveals evidence of electron accumulation in the solar photosphere. This is not the case of the heavier protons, which remain in lower layers. An electric field would thus be present in the solar interior, but as the total charge remains negligible, no electric field or effect would result outside the star.}


\offprints{V. Bommier, \email{V.Bommier@obspm.fr}}

\maketitle

\titlerunning{Sun's photosphere negatively charged ?}
\authorrunning{V. Bommier}

\section{Introduction}

The problem of the large magnitude difference between the observed horizontal and vertical magnetic field gradients in and around sunspots has been known for a long time. \citet{Balthasar-18} wrote a detailed review of observations, where it is shown that  typical values of 3 G/km and 0.3 G/km are obtained for the vertical and horizontal gradients of the magnetic field, respectively, regardless of the telescope, spectral line(s), and measurement interpretation method  used. Using these values would surprisingly lead to a non-zero divergence of the observed magnetic field, which is a priori not acceptable. In his review ``Sunspots: An overview,'' \citet[p. 184]{Solanki-03} expressed the problem as follows: No satisfactory solution has been found as yet for the unexpectedly small vertical gradients obtained by applying the $\mathrm{div}\vec{B}=0$ condition [to the observed horizontal gradients]. 

\citet{Balthasar-18} tries to explain the discrepancy by simulating unresolved magnetic structures. This does not manage to fully explain the observations. In Sect. \ref{sect--inacc} we rule out the effect of measurement inaccuracies. In Sect. \ref{sect--maths}, we rule out a pure effect of spatial resolution on the basis of two mathematical theorems. We discuss the difference between derivatives and finite differences (the observations), and we show that these mathematical theorems prove that the observed non-zero divergence reveals the existence of at least one non-zero contribution from an element of the averaged region. 

In this paper we   explain the observations by applying the Maxwell relation in magnetized media $\vec{B}=\mu_0\left(\vec{H}+\vec{M}\right)$, provided that it can be proved that what is measured by the Zeeman effect is $\vec{H}$ and not $\vec{B}$. This is the object of Sect. \ref{sect--Zeeman}. To establish this point it is necessary to go back to the microscopic scale of an atom embedded in the medium magnetized at the macroscopic scale. In addition, we study the atom potential magnetic energy. As introduced in Sect. \ref{sect--fields}, in this paper we denote as $\vec{B}$ the magnetic induction and as $\vec{H}$ the magnetic field, which are related by the law 
$\vec{B}=\mu_0\left(\vec{H}+\vec{M}\right)$, where $\vec{M}$ is the magnetization.

However, the photosphere magnetization is negligible when evaluated via the usual models. In Sect. \ref{sect--trap} we propose a model that is able to increase magnetization in the photosphere. It is considered that in the solar interior, at 0.5 $R_{\sun}$, the electron thermal velocity of 12 Mm/s largely surpasses the escape velocity of 850 km/s; however,  this is not the case for  protons whose thermal velocity is 290 km/s, only due to their much higher mass. A similar effect occurs in the Solar Corona \citep{Meyer-Vernet-07}. Following \citet{Allen-73}, the mass inside the sphere of 0.5 $R_{\sun}$ radius is 0.94 of the total solar mass, and the temperature is $3.4 \times 10^6$ K there. This leads to the above velocity values. Thus, gravity separates the charges, in the sense that electrons escape when protons do not. More precisely, the underlying protons do not completely prevent electrons from escaping. The result is a free electron accumulation and a space charge at the Sun's surface. We show that the electron migration is a quasi-static process. As a result, the photosphere magnetization due to the plasma diamagnetism, which is  itself due to these charges, could become non-negligible. Under the usual photospheric conditions \citep[the VALC model;][]{Vernazza-etal-81}, the magnetization would be about $10^{-4}$ lower than $B/\mu_0$, due to the low electron and ion densities \citep{Bommier-15}. The accumulation of escaping electrons could increase the magnetization up to a non-negligible level.

In Sect. \ref{sect--measurements} we present the most direct electron density measurements in the solar photosphere to our knowledge. We propose that these measurements show evidence of an electron overdensity in the solar photosphere. We conclude in Sect. \ref{sect--concl}.

\section{Measurement inaccuracies}
\label{sect--inacc}

\begin{figure*}
{\includegraphics[width=6.16cm]{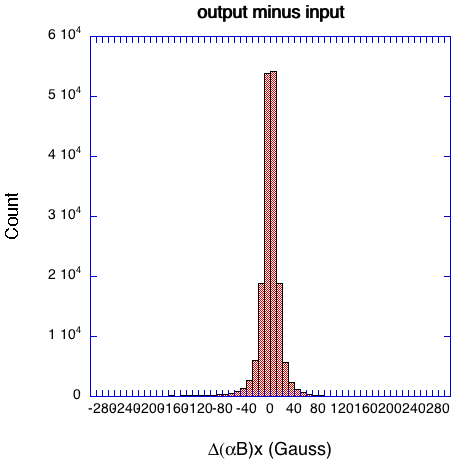}}{\includegraphics[width=6.16cm]{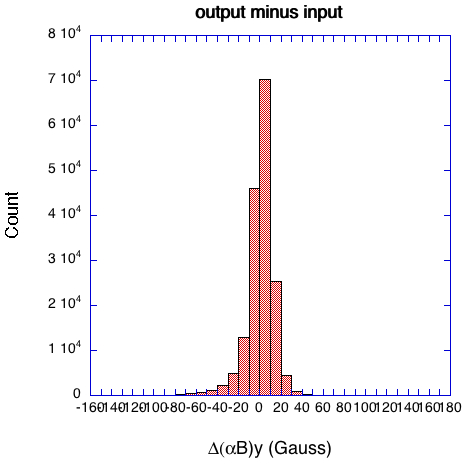}}{\includegraphics[width=6.16cm]{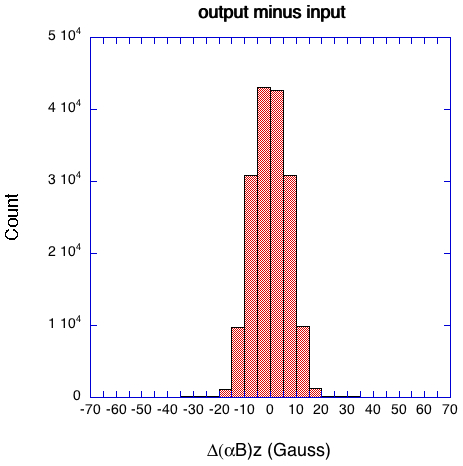}}
\caption{Histograms of inversion results of noised theoretical polarization profiles. The introduced noise level is $1.5 \times 10^{-3}$, in terms of polarization degree. The  method used is a Milne-Eddington inversion, but including a magnetic filling factor $\alpha$ (see text). As a result, the inaccuracy can be estimated to be 20 G for the $x$ and $y$ components (left  and middle plots), whereas it is 10 G for the $z$ component (right plot). The $Oz$-axis lies along the line of sight. The assumed line is \ion{Fe}{i} 6302.5 \AA. These histograms were published (in spherical coordinates) in Fig. 5 of \citet{Bommier-etal-07}.}
\label{histos}
\end{figure*}

\citet{Balthasar-18} published a review of observations of the magnetic field gradients in and around sunspots. The measurement inaccuracy obviously depends on the method and  the instrument used. However, we find it interesting to publish here the histograms of magnetic inaccuracy obtained within the UNNOFIT inversion method that we recently developed \citep{Bommier-etal-07}, in magnetic field cartesian coordinates. These histograms can be found in Fig. 5 of the above-mentioned paper, but in spherical coordinates, which prevents us from clearly discriminating between longitudinal and transverse field components. In Fig. \ref{histos} of the present paper, we publish the histograms for each of these components. As described in \citet{Bommier-etal-07}, these histograms were obtained from inversion of theoretical noised polarization profiles. The noise level was assumed to be $1.5 \times 10^{-3}$, in terms of polarization degree. The magnetic field strength ranged from 100 to 3000 G. The inversion method is of the Milne-Eddington type, but assuming a two-component atmosphere represented with a magnetic filling factor $\alpha$. The originality of our method lies in the fact that the magnetic filling factor is entered as the ninth free parameter in the Levenberg-Marquardt algorithm.

Figure \ref{histos} shows that the magnetic inaccuracy is on the order of 20 G for the transverse component, while  it is on the order of 10 G for the longitudinal component. We applied our method to observations of the \ion{Fe}{i} 6301.5 and 6302.5 \AA\ lines, which are formed with a height difference of about 100 km in active regions, as shown in Fig. 4 of \citet{Khomenko-Collados-07}. As the field strength  there is on the order of one thousand gauss, with a difference of about 300 G between the two line measurements, leading to the vertical gradient value of 3 G/km,  as reported by \citet{Balthasar-18}, the 10 G inaccuracy on the longitudinal field is much smaller than the finite difference of the measurements. The horizontal gradient  is found to be only about 0.3 G/km, which corresponds to the reversal of a 1500 G typical horizontal component from one side to the other  of a sunspot with a  typical diameter of 10,000 km. Assuming a 0.3 arcsec pixel, which is 250 km wide, the observed horizontal field variation is  about 75 G, which is not very large, yet larger than the 20 G inaccuracy on the transverse field. As a result, the difference between the observed vertical and horizontal gradients cannot be simply assigned to measurement inaccuracies.

\section{Some mathematical theorems}
\label{sect--maths}

In this section we determine whether the lack of spatial resolution is able to explain the observed discrepancy. First of all we note that if the lack of spatial resolution were able to explain the effect, we would expect   the non-zero value observed for the divergence   to depend on the pixel size. As far   can be seen in the review by \citet{Balthasar-18}, this is not the case. The review deals with various instruments, and therefore various pixel sizes, but the results are   homogeneous, even though obtained   using different active regions. In order to investigate this question in a more quantitative manner, we considered the spectropolarimetric data of NOAA 10953 acquired by the SOT/SP experiment on board the HINODE satellite on 1 May 2007 at 10:46 UT. The pixel size was $0.16\arcsec$. We then prepared artificially degraded resolution data by averaging the Stokes parameters on $2^2$, $4^2$, and $8^2$ pixels. We then submitted these artificially degraded resolution data together with the original data to our inversion code UNNOFIT \citep{Bommier-etal-07}. We did not detect any significant variation in the difference between vertical and horizontal magnetic field gradients as a function of the pixel size.

It can then be objected that the typical variation length for the magnetic field may be very small with respect to these pixel sizes (or typical pixel sizes or current instrument pixel sizes). In this case no variation with the pixel size could be observed because of the scale difference between the pixel size and the magnetic field typical variation length. However, such a case can be treated by the theorems we introduce below. 

\subsection{Theorem 1: filtering and derivation commute}

We define  the filtered quantity through a filter not necessarily isotropic as a local average. The filtering operation is a convolution product of the quantity with the filter function. Theorem 1 is that this filtering (or convolution) operation commutes with the derivation operation. The result is that the derivative or divergence of the local average is the local average of the derivatives or divergences respectively. We recall below our demonstration, presented in \citet{Bommier-13,Bommier-14}.

Due to the linearity of the Zeeman effect, as previously discussed, the line-of-sight integration can be modeled as
\begin{equation}
% MathType!MTEF!2!1!+-
% faaahqart1ev3aaaKnaaaaWenf2ys9wBH5garuavP1wzZbItLDhis9
% wBH5garmWu51MyVXgaruWqVvNCPvMCaebbnrfifHhDYfgasaacH8sr
% ps0lbbf9q8WrFfeuY-ribbf9v8qqaqFr0xc9pk0xbba9q8WqFfea0-
% yr0RYxir-Jbba9q8aq0-yq-He9q8qqQ8frFve9Fve9Ff0dc9Gqpi0d
% meaabaqaciGacaGaaeqabaWaaeaaeaaakeaaqaaaaaaaaaWdbiqadI
% eagaWcamaabmaabaGaamOEaaGaayjkaiaawMcaaiabg2da9maapeda
% paqaa8qaceWGibGbaSaadaqadaqaaiqadQhagaqbaaGaayjkaiaawM
% caaiabeA8aQnaabmaabaGaamOEaiabgkHiTiqadQhagaqbaaGaayjk
% aiaawMcaaiGacsgaaSWdaeaapeGaeyOeI0IaeyOhIukapaqaa8qacq
% GHRaWkcqGHEisPa0Gaey4kIipakiqadQhagaqbaaaa!48F2!
\vec H\left( z \right) = \int_{ - \infty }^{ + \infty } {\vec H\left( {z'} \right)\varphi \left( {z - z'} \right){\mathop{\rm d}\nolimits} } z'  \ \ ,
\end{equation}
where $\varphi$ is the contribution function, and where we have assigned a ``height of formation'' $z$ to the final result. This height of formation is generally close to the maximum of the contribution function, as seen in the examples computed by \citet{Bruls-etal-91}. The contribution function acts as a filter in the above equation. Analogously, the pixel integration in $x$ or $y$ can be modeled with a convolution by a crenel function.

If now we compute the divergence of the observed magnetic field, we can apply the derivation of a convolution product as recalled below. Given the convolution product defined as
\begin{equation}
% MathType!MTEF!2!1!+-
% faaahqart1ev3aaaKnaaaaWenf2ys9wBH5garuavP1wzZbItLDhis9
% wBH5garmWu51MyVXgaruWqVvNCPvMCaebbnrfifHhDYfgasaacH8sr
% ps0lbbf9q8WrFfeuY-ribbf9v8qqaqFr0xc9pk0xbba9q8WqFfea0-
% yr0RYxir-Jbba9q8aq0-yq-He9q8qqQ8frFve9Fve9Ff0dc9Gqpi0d
% meaabaqaciGacaGaaeqabaWaaeaaeaaakeaaqaaaaaaaaaWdbiaadA
% eadaqadaqaaiaadIhaaiaawIcacaGLPaaacqGH9aqpdaWdXaWdaeaa
% peGaamOzamaabmaabaGabmiEayaafaaacaGLOaGaayzkaaGaam4zam
% aabmaabaGaamiEaiabgkHiTiqadIhagaqbaaGaayjkaiaawMcaaiGa
% csgaaSWdaeaacaWGbbaabaGaamOqaaqdpeGaey4kIipakiqadIhaga
% qbaaaa!44CC!
F\left( x \right) = \int_A^B {f\left( {x'} \right)g\left( {x - x'} \right){\mathop{\rm d}\nolimits} } x'  \ \ ,
\end{equation}
we evaluate 
% MathType!MTEF!2!1!+-
% faaahqart1ev3aaaKnaaaaWenf2ys9wBH5garuavP1wzZbItLDhis9
% wBH5garmWu51MyVXgaruWqVvNCPvMCaebbnrfifHhDYfgasaacH8sr
% ps0lbbf9q8WrFfeuY-ribbf9v8qqaqFr0xc9pk0xbba9q8WqFfea0-
% yr0RYxir-Jbba9q8aq0-yq-He9q8qqQ8frFve9Fve9Ff0dc9Gqpi0d
% meaabaqaciGacaGaaeqabaWaaeaaeaaakeaacaWGgbWaaeWaaeaaca
% WG4bGaey4kaSIaciizaiaadIhaaiaawIcacaGLPaaaaaa!3799!
$F\left( {x + {\mathop{\rm d}\nolimits} x} \right)$ 
in order to evaluate the derivative. We obtain 
\begin{eqnarray}
% MathType!MTEF!2!1!+-
% faaahqart1ev3aaaKnaaaaWenf2ys9wBH5garuavP1wzZbItLDhis9
% wBH5garmWu51MyVXgaruWqVvNCPvMCaebbnrfifHhDYfgasaacH8sr
% ps0lbbf9q8WrFfeuY-ribbf9v8qqaqFr0xc9pk0xbba9q8WqFfea0-
% yr0RYxir-Jbba9q8aq0-yq-He9q8qqQ8frFve9Fve9Ff0dc9Gqpi0d
% meaabaqaciGacaGaaeqabaWaaeaaeaaakqaabeqaaiaadAeadaqada
% qaaiaadIhacqGHRaWkciGGKbGaamiEaaGaayjkaiaawMcaaabaaaaa
% aaaapeGaeyypa0Zaa8qma8aabaWdbiaadAgadaqadaqaaiqadIhaga
% qbaaGaayjkaiaawMcaaiaadEgadaqadaqaaiaadIhapaGaey4kaSIa
% ciizaiaadIhapeGaeyOeI0IabmiEayaafaaacaGLOaGaayzkaaGaci
% izaaWcpaqaaiaadgeacqGHRaWkciGGKbGaamiEaaqaaiaadkeacqGH
% RaWkciGGKbGaamiEaaqdpeGaey4kIipakiqadIhagaqbaaqaaiabg2
% da9maapedapaqaa8qacaWGMbWaaeWaaeaaceWG4bGbayaapaGaey4k
% aSIaciizaiaadIhaa8qacaGLOaGaayzkaaGaam4zamaabmaabaGaam
% iEaiabgkHiTiqadIhagaGbaaGaayjkaiaawMcaaiGacsgaaSWdaeaa
% caWGbbaabaGaamOqaaqdpeGaey4kIipakiqadIhagaGbaaaaaa!62DD!
F\left( {x + {\mathop{\rm d}\nolimits} x} \right) = \int_{A + {\mathop{\rm d}\nolimits} x}^{B + {\mathop{\rm d}\nolimits} x} {f\left( {x'} \right)g\left( {x + {\mathop{\rm d}\nolimits} x - x'} \right){\mathop{\rm d}\nolimits} } x'\\
 = \int_A^B {f\left( {x'' + {\mathop{\rm d}\nolimits} x} \right)g\left( {x - x''} \right){\mathop{\rm d}\nolimits} } x''  \ \ ,
\end{eqnarray}
by changing the variable 
% MathType!MTEF!2!1!+-
% faaahqart1ev3aaaKnaaaaWenf2ys9wBH5garuavP1wzZbItLDhis9
% wBH5garmWu51MyVXgaruWqVvNCPvMCaebbnrfifHhDYfgasaacH8sr
% ps0lbbf9q8WrFfeuY-ribbf9v8qqaqFr0xc9pk0xbba9q8WqFfea0-
% yr0RYxir-Jbba9q8aq0-yq-He9q8qqQ8frFve9Fve9Ff0dc9Gqpi0d
% meaabaqaciGacaGaaeqabaWaaeaaeaaakeaaceWG4bGbayaacqGH9a
% qpceWG4bGbauaacqGHsislciGGKbGaamiEaaaa!376C!
$x'' = x' - {\mathop{\rm d}\nolimits} x$.
We then have   the derivative 
% MathType!MTEF!2!1!+-
% faaahqart1ev3aaaKnaaaaWenf2ys9wBH5garuavP1wzZbItLDhis9
% wBH5garmWu51MyVXgaruWqVvNCPvMCaebbnrfifHhDYfgasaacH8sr
% ps0lbbf9q8WrFfeuY-ribbf9v8qqaqFr0xc9pk0xbba9q8WqFfea0-
% yr0RYxir-Jbba9q8aq0-yq-He9q8qqQ8frFve9Fve9Ff0dc9Gqpi0d
% meaabaqaciGacaGaaeqabaWaaeaaeaaakeaaqaaaaaaaaaWdbiqadA
% eagaqbamaabmaabaGaamiEaaGaayjkaiaawMcaaaaa!34FC!
$F'\left( x \right)$
\begin{equation}
% MathType!MTEF!2!1!+-
% faaahqart1ev3aaaKnaaaaWenf2ys9wBH5garuavP1wzZbItLDhis9
% wBH5garmWu51MyVXgaruWqVvNCPvMCaebbnrfifHhDYfgasaacH8sr
% ps0lbbf9q8WrFfeuY-ribbf9v8qqaqFr0xc9pk0xbba9q8WqFfea0-
% yr0RYxir-Jbba9q8aq0-yq-He9q8qqQ8frFve9Fve9Ff0dc9Gqpi0d
% meaabaqaciGacaGaaeqabaWaaeaaeaaakeaaqaaaaaaaaaWdbiqadA
% eagaqbamaabmaabaGaamiEaaGaayjkaiaawMcaaiabg2da9maapeda
% paqaa8qaceWGMbGbauaadaqadaqaaiqadIhagaqbaaGaayjkaiaawM
% caaiaadEgadaqadaqaaiaadIhacqGHsislceWG4bGbauaaaiaawIca
% caGLPaaaciGGKbaal8aabaGaamyqaaqaaiaadkeaa0WdbiabgUIiYd
% GcceWG4bGbauaaaaa!44E4!
F'\left( x \right) = \int_A^B {f'\left( {x'} \right)g\left( {x - x'} \right){\mathop{\rm d}\nolimits} } x'  \ \ ,
\end{equation}
where  the derivative of the convolution product is the convolution (by the same function) of the derivative.

Returning to the case of the line-of-sight integration for the divergence, which is a combination of derivatives:
\begin{equation}
% MathType!MTEF!2!1!+-
% faaahqart1ev3aaaKnaaaaWenf2ys9wBH5garuavP1wzZbItLDhis9
% wBH5garmWu51MyVXgaruWqVvNCPvMCaebbnrfifHhDYfgasaacH8sr
% ps0lbbf9q8WrFfeuY-ribbf9v8qqaqFr0xc9pk0xbba9q8WqFfea0-
% yr0RYxir-Jbba9q8aq0-yq-He9q8qqQ8frFve9Fve9Ff0dc9Gqpi0d
% meaabaqaciGacaGaaeqabaWaaeaaeaaakeaaciGGKbGaaiyAaiaacA
% haceWGibGbaSaadaqadaqaaiaadIhacaGGSaGaamyEaiaacYcacaWG
% 6baacaGLOaGaayzkaaGaeyypa0Zaa8qmaeaaciGGKbGaaiyAaiaacA
% haceWGibGbaSaadaqadaqaaiaadIhacaGGSaGaamyEaiaacYcaceWG
% 6bGbauaaaiaawIcacaGLPaaaaSqaaiabgkHiTiabg6HiLcqaaiabgU
% caRiabg6HiLcqdcqGHRiI8aOGaeqOXdO2aaeWaaeaacaWG6bGaeyOe
% I0IabmOEayaafaaacaGLOaGaayzkaaGaciizaiqadQhagaqbaaaa!54CD!
{\mathop{\rm div}\nolimits} \vec H\left( {x,y,z} \right) = \int_{ - \infty }^{ + \infty } {{\mathop{\rm div}\nolimits} \vec H\left( {x,y,z'} \right)} \varphi \left( {z - z'} \right){\mathop{\rm d}\nolimits} z'  \ \ .
\end{equation}
An analogous derivation can be made for the filtering with a crenel function for the pixel integration. It results from the derivation property of the convolution product that any filter would lead to the same result, which is that the divergence of the filtered quantity is the filtering applied to the divergence of the local quantity.

This result can be more easily derived in the Fourier space, where convolution products are transformed into simple products. If we denote as 
% MathType!MTEF!2!1!+-
% faaahqart1ev3aaaKnaaaaWenf2ys9wBH5garuavP1wzZbItLDhis9
% wBH5garmWu51MyVXgaruWqVvNCPvMCaebbnrfifHhDYfgasaacH8sr
% ps0lbbf9q8WrFfeuY-ribbf9v8qqaqFr0xc9pk0xbba9q8WqFfea0-
% yr0RYxir-Jbba9q8aq0-yq-He9q8qqQ8frFve9Fve9Ff0dc9Gqpi0d
% meaabaqaciGacaGaaeqabaWaaeaaeaaakeaaceWGibGbaKaadaWgaa
% WcbaGaamiEaaqabaGcdaqadaqaaiqadUgagaWcaaGaayjkaiaawMca
% aaaa!361A!
${\hat H_x}\left( {\vec k} \right)$
(resp. $y$, $z$) the spatial Fourier transform of the magnetic field component 
% MathType!MTEF!2!1!+-
% faaahqart1ev3aaaKnaaaaWenf2ys9wBH5garuavP1wzZbItLDhis9
% wBH5garmWu51MyVXgaruWqVvNCPvMCaebbnrfifHhDYfgasaacH8sr
% ps0lbbf9q8WrFfeuY-ribbf9v8qqaqFr0xc9pk0xbba9q8WqFfea0-
% yr0RYxir-Jbba9q8aq0-yq-He9q8qqQ8frFve9Fve9Ff0dc9Gqpi0d
% meaabaqaciGacaGaaeqabaWaaeaaeaaakeaacaWGibWaaSbaaSqaai
% aadIhaaeqaaOWaaeWaaeaaceWGYbGbaSaaaiaawIcacaGLPaaaaaa!3611!
${H_x}\left( {\vec r} \right)$
(resp. $y$, $z$), the Fourier transform of the field divergence is
\begin{equation}
% MathType!MTEF!2!1!+-
% faaahqart1ev3aaaKnaaaaWenf2ys9wBH5garuavP1wzZbItLDhis9
% wBH5garmWu51MyVXgaruWqVvNCPvMCaebbnrfifHhDYfgasaacH8sr
% ps0lbbf9q8WrFfeuY-ribbf9v8qqaqFr0xc9pk0xbba9q8WqFfea0-
% yr0RYxir-Jbba9q8aq0-yq-He9q8qqQ8frFve9Fve9Ff0dc9Gqpi0d
% meaabaqaciGacaGaaeqabaWaaeaaeaaakeaacaWGgbGaamivamaadm
% aabaGaciizaiaacMgacaGG2bGabmisayaalaaacaGLBbGaayzxaaGa
% eyypa0JaciyAaiaadUgadaWgaaWcbaGaamiEaaqabaGcceWGibGbaK
% aadaWgaaWcbaGaamiEaaqabaGccqGHRaWkciGGPbGaam4AamaaBaaa
% leaacaWG5baabeaakiqadIeagaqcamaaBaaaleaacaWG5baabeaaki
% abgUcaRiGacMgacaWGRbWaaSbaaSqaaiaadQhaaeqaaOGabmisayaa
% jaWaaSbaaSqaaiaadQhaaeqaaaaa!4AF1!
FT\left[ {{\mathop{\rm div}\nolimits} \vec H} \right] = {\mathop{\rm i}\nolimits} {k_x}{\hat H_x} + {\mathop{\rm i}\nolimits} {k_y}{\hat H_y} + {\mathop{\rm i}\nolimits} {k_z}{\hat H_z} \ \ .
\end{equation}
We denote as 
% MathType!MTEF!2!1!+-
% faaahqart1ev3aaaKnaaaaWenf2ys9wBH5garuavP1wzZbItLDhis9
% wBH5garmWu51MyVXgaruWqVvNCPvMCaebbnrfifHhDYfgasaacH8sr
% ps0lbbf9q8WrFfeuY-ribbf9v8qqaqFr0xc9pk0xbba9q8WqFfea0-
% yr0RYxir-Jbba9q8aq0-yq-He9q8qqQ8frFve9Fve9Ff0dc9Gqpi0d
% meaabaqaciGacaGaaeqabaWaaeaaeaaakeaacqaHgpGAdaqadaqaai
% qadkhagaWcaaGaayjkaiaawMcaaaaa!35CE!
$\varphi \left( {\vec r} \right)$ 
the 3D spatial filter to be applied to model the observations, and we accordingly denote as
% MathType!MTEF!2!1!+-
% faaahqart1ev3aaaKnaaaaWenf2ys9wBH5garuavP1wzZbItLDhis9
% wBH5garmWu51MyVXgaruWqVvNCPvMCaebbnrfifHhDYfgasaacH8sr
% ps0lbbf9q8WrFfeuY-ribbf9v8qqaqFr0xc9pk0xbba9q8WqFfea0-
% yr0RYxir-Jbba9q8aq0-yq-He9q8qqQ8frFve9Fve9Ff0dc9Gqpi0d
% meaabaqaciGacaGaaeqabaWaaeaaeaaakeaacuaHgpGAgaqcamaabm
% aabaGabm4AayaalaaacaGLOaGaayzkaaaaaa!35D7!
$\hat \varphi \left( {\vec k} \right)$ 
its Fourier transform. Because
\begin{equation}
% MathType!MTEF!2!1!+-
% MathType!MTEF!2!1!+-
% faaahqart1ev3aaaKnaaaaWenf2ys9wBH5garuavP1wzZbItLDhis9
% wBH5garmWu51MyVXgaruWqVvNCPvMCaebbnrfifHhDYfgasaacH8sr
% ps0lbbf9q8WrFfeuY-ribbf9v8qqaqFr0xc9pk0xbba9q8WqFfea0-
% yr0RYxir-Jbba9q8aq0-yq-He9q8qqQ8frFve9Fve9Ff0dc9Gqpi0d
% meaabaqaciGacaGaaeqabaWaaeaaeaaakeaacuaHgpGAgaqcaiabgw
% SixpaadmaabaGaciyAaiaadUgadaWgaaWcbaGaamiEaaqabaGcceWG
% ibGbaKaadaWgaaWcbaGaamiEaaqabaGccqGHRaWkciGGPbGaam4Aam
% aaBaaaleaacaWG5baabeaakiqadIeagaqcamaaBaaaleaacaWG5baa
% beaakiabgUcaRiGacMgacaWGRbWaaSbaaSqaaiaadQhaaeqaaOGabm
% isayaajaWaaSbaaSqaaiaadQhaaeqaaaGccaGLBbGaayzxaaGaeyyp
% a0JaciyAaiaadUgadaWgaaWcbaGaamiEaaqabaGccuaHgpGAgaqcai
% qadIeagaqcamaaBaaaleaacaWG4baabeaakiabgUcaRiGacMgacaWG
% RbWaaSbaaSqaaiaadMhaaeqaaOGafqOXdOMbaKaaceWGibGbaKaada
% WgaaWcbaGaamyEaaqabaGccqGHRaWkciGGPbGaam4AamaaBaaaleaa
% caWG6baabeaakiqbeA8aQzaajaGabmisayaajaWaaSbaaSqaaiaadQ
% haaeqaaaaa!604B!
\hat \varphi  \cdot \left[ {{\mathop{\rm i}\nolimits} {k_x}{{\hat H}_x} + {\mathop{\rm i}\nolimits} {k_y}{{\hat H}_y} + {\mathop{\rm i}\nolimits} {k_z}{{\hat H}_z}} \right] = {\mathop{\rm i}\nolimits} {k_x}\hat \varphi {\hat H_x} + {\mathop{\rm i}\nolimits} {k_y}\hat \varphi {\hat H_y} + {\mathop{\rm i}\nolimits} {k_z}\hat \varphi {\hat H_z} \ \ ,
\end{equation}
we have 
% MathType!MTEF!2!1!+-
% faaahqart1ev3aaaKnaaaaWenf2ys9wBH5garuavP1wzZbItLDhis9
% wBH5garmWu51MyVXgaruWqVvNCPvMCaebbnrfifHhDYfgasaacH8sr
% ps0lbbf9q8WrFfeuY-ribbf9v8qqaqFr0xc9pk0xbba9q8WqFfea0-
% yr0RYxir-Jbba9q8aq0-yq-He9q8qqQ8frFve9Fve9Ff0dc9Gqpi0d
% meaabaqaciGacaGaaeqabaWaaeaaeaaakeaacqaHgpGAdaWadaqaai
% GacsgacaGGPbGaaiODaiqadIeagaWcaaGaay5waiaaw2faaiabg2da
% 9iGacsgacaGGPbGaaiODaiabeA8aQnaadmaabaGabmisayaalaaaca
% GLBbGaayzxaaaaaa!4143!
$\varphi \left[ {{\mathop{\rm div}\nolimits} \vec H} \right] = {\mathop{\rm div}\nolimits} \varphi \left[ {\vec H} \right]$,
which is the above-mentioned result. This does not assume any isotropy of the filter. Different filter sizes and types may be assumed in $x$, $y$, $z$, as they are in  the case of the observations, and the result will be maintained.

\subsection{Derivatives and finite differences}

In practice, the divergence of the magnetic field vector in the observations is evaluated by means of finite differences and not mathematical derivatives. In order to compare derivatives and finite differences, we have to discriminate between typical lengths of field variation shorter than, similar to, or longer than the  typical pixel size used as the typical length for local averaging.

The field variations shorter than the pixel size may have local derivatives larger or smaller than the derivative of the averaged function, which is also the average of the derivatives over the pixel. The averaged derivative over the pixel corresponds to larger variations with respect to the pixel size. By integrating over the pixel, the average function variations yield typical variation lengths that are longer than the pixel size. Only those remain, but the smaller values are included in the average.

When the typical variation lengths are longer than the pixel size, function derivatives and finite differences at the pixel size are close together. The derivatives computed by finite differences at the pixel size are then very close to the averaged derivatives over the pixel.

The case of typical variation lengths similar to the pixel size is a limit case. In this case, the effect of a difference between derivative and finite difference, would be sensitive to the pixel size. This is not what is observed, as discussed at the beginning of this section.

The typical height difference between the observed lines is on the order of 63 km in the quiet photosphere, as directly observed by \citet{Faurobert-etal-09} for the two usual lines,  \ion{Fe}{i} 6301.5 and 6302.5 \AA, and slightly more in active regions following the simulation by \citet[][Fig. 4]{Khomenko-Collados-07}. We also note that  when such a height difference is combined with the observed vertical gradient of 3 G/km, this leads to a field decrease of about 250 G in this interval, as observed, when the field itself is about 2000 G. Therefore, the $\Delta z$ used to determine the vertical gradient is significantly smaller than the magnetic field strength height scale, which is the height difference for the magnetic field strength being divided by two, which increases the significance of the result.

For the observed horizontal gradient, its order of magnitude of 0.3 G/km is fully compatible with the typical sunspot diameter (10,000 km) and the typical horizontal field component in the penumbra (1500 G), which reverses from one side of the sunspot to the opposite side. It is probably not underestimated.

Alternatively, if it is assumed that the vertical gradient is overestimated, and if consequently it were instead  ten times smaller (on the order of 0.3 G/km also), the height formation difference between the two lines would then be ten times larger, from 700 to 1000 km, which is inconsistent with the photosphere visible thickness.

\subsection{Theorem 2: the content of the non-zero average}

When the averaged quantities are all zero, the average is accordingly zero. By taking the opposite of this statement, we obtain the second theorem: if the averaged quantity is non-zero, there is at least one quantity among the averaged quantities that is non-zero.

Applying this second theorem to our problem, we obtain that if a non-zero divergence of the magnetic field is obtained  from the averaged quantities, there is at least one point within the pixel where the divergence is non-zero. A non-zero divergence cannot result  from the lack of spatial resolution alone because it is made up of averaging operations. The observed divergence is probably true.

It should  be recalled that the Zeeman effect, which is responsible for the effect of the magnetic field on the spectral lines, is an intrinsically linear effect because the sublevel energy variation is linear as a function of the magnetic field strength. 

\section{Magnetic induction, magnetic field, and magnetization}
\label{sect--fields}

We recall that the three quantities of magnetic induction,  magnetic field, and  magnetization are related by the Maxwell equation in magnetized media
\begin{equation}
\vec{B}=\mu_0\left(\vec{H}+\vec{M}\right) \ \ ,
\label{eq--BHM}
\end{equation}
where $\vec{B}$ is the magnetic induction, $\vec{H}$ is the magnetic field, and $\vec{M}$ the magnetization. 

For the definition of these quantities, we  refer to \citet{Jackson-75}, in particular Sect. 6.7, ``Derivation of the Equations of Macroscopic Electromagnetism,'' where Eq. (\ref{eq--BHM}) is derived by averaging from the microscopic scale, with an averaging length that is large with respect to the microscopic scale, but small with respect to the macroscopic scale. Equation (\ref{eq--BHM}) is then macroscopic. At the microscopic scale, the electric and magnetic fields are denoted  $\vec{e}$ and $\vec{b}$, respectively, in lowercase characters. They obey the Maxwell equations in a vacuum, where the charge and current densities include all the charges present in the medium. By separating the effects of free and bound charges, the four macroscopic quantities $\vec{D}$, $\vec{H}$, $\vec{P}$, and $\vec{M}$ are found at the macroscopic scale, where the polarization  $\vec{P}$ and the magnetization $\vec{M}$ are due to the contribution of the  bound charges. \citet{Jackson-75} tells us (p. 233) that when the medium is a plasma that also contains free charges, and if these free charges also possess intrinsic magnetic moments, these magnetic moments can be simply included in the definition of $\vec{M}$.

\citet[vol. 1, p 89]{Delcroix-Bers-94} evaluate this intrinsic magnetic moment of the plasma free charges in their spiral motion about the magnetic induction due to the Lorentz force $q\vec{v} \times \vec{B}$. They show  for the ensemble of orbital magnetic moments of the individual particles 
\begin{equation}
% MathType!MTEF!2!1!+-
% faaahqart1ev3aaaKnaaaaWenf2ys9wBH5garuavP1wzZbItLDhis9
% wBH5garmWu51MyVXgaruWqVvNCPvMCaebbnrfifHhDYfgasaacH8sr
% ps0lbbf9q8WrFfeuY-ribbf9v8qqaqFr0xc9pk0xbba9q8WqFfea0-
% yr0RYxir-Jbba9q8aq0-yq-He9q8qqQ8frFve9Fve9Ff0dc9Gqpi0d
% meaabaqaciGacaGaaeqabaWaaeaaeaaakeaaceWGnbGbaSaacqGH9a
% qpcqGHsisldaWcaaqaaiabek7aIbqaaiaaikdacqaH8oqBdaWgaaWc
% baGaaGimaaqabaaaaOGabmOqayaalaaaaa!3A42!
\vec M =  - \frac{\beta }{{2{\mu _0}}}\vec B \ \ ,
\label{eq--diamagnetism}
\end{equation}
where $\beta$ is the plasma usual parameter, but computed with the free charge density. This magnetic moment has the opposite sign with respect to the magnetic induction. The effect is therefore called plasma diamagnetism.

The photospheric VALC model \citep{Vernazza-etal-81} was derived from spectroscopic observations, statistical equilibrium, radiative transfer, and the Saha law for deriving the electron density within the hypothesis of medium electric neutrality. At the formation height of the \ion{Fe}{i} lines used for the magnetic field diagnostic (about 250 km) the temperature is 4780 K, the neutral hydrogen density is $2.3 \times 10^{16}$, and the electron density is $2.7 \times 10^{12}$ cm$^{-3}$. The plasma $\beta$ expressed in terms of the electron density is $\beta = 4.4 \times 10^{-5}$ (for $B=1000$ G) and $\vec{M}$ is negligible with respect to $\vec{H}$ and $\vec{B}/\mu_0$. The main second source of magnetization is the neutral hydrogen  paramagnetism. Its susceptibility is $\chi=3 \times 10^{-5}$ at 250 km (correcting  an error  present in \citealt{Bommier-13,Bommier-14}), with $\vec{M} = \chi \vec{H}$. The neutral  iron paramagnetism susceptibility is weaker $\chi=2 \times 10^{-8}$. We  have in this case $\vec{B} \simeq \mu_0 \vec{H}$,  but if there were free electrons coming ``from below'' and accumulating in the photosphere as we propose, and if their density were finally  comparable to the neutral hydrogen density, we would have $\beta = 0.38$ (for $B=1000$ G). $\vec{M}$ would then be comparable to $\vec{H}$ and $\vec{B}/\mu_0$, and they would  all be different.

In the Maltby-M sunspot umbra model \citep{Maltby-etal-86}, the formation height of the \ion{Fe}{i} lines is about 100 km. The $\beta$ expressed in terms of the neutral hydrogen density is $\beta=1.27$ at this height and for $B=1000$ G, which is $\beta=0.20$ for the magnetic field $B=2500$ G typical of sunspot umbra. 

We assume for a moment an oversimplified model of sunspot umbra with a purely vertical field or induction along $z$. The Mawxell law $\mathrm{div}\vec{B}=0$ then results in 
% MathType!MTEF!2!1!+-
% faaahqart1ev3aaaKnaaaaWenf2ys9wBH5garuavP1wzZbItLDhis9
% wBH5garmWu51MyVXgaruWqVvNCPvMCaebbnrfifHhDYfgasaacH8sr
% ps0lbbf9q8WrFfeuY-ribbf9v8qqaqFr0xc9pk0xbba9q8WqFfea0-
% yr0RYxir-Jbba9q8aq0-yq-He9q8qqQ8frFve9Fve9Ff0dc9Gqpi0d
% meaabaqaciGacaGaaeqabaWaaeaaeaaakeaadaWcgaqaaiabgkGi2k
% aadkeadaWgaaWcbaGaamOEaaqabaaakeaacqGHciITcaWG6baaaiab
% g2da9iaaicdaaaa!391C!
${{\partial {B_z}} \mathord{\left/
 {\vphantom {{\partial {B_z}} {\partial z}}} \right.
 \kern-\nulldelimiterspace} {\partial z}} = 0$.
We then have
\begin{equation}
% MathType!MTEF!2!1!+-
% faaahqart1ev3aaaKnaaaaWenf2ys9wBH5garuavP1wzZbItLDhis9
% wBH5garmWu51MyVXgaruWqVvNCPvMCaebbnrfifHhDYfgasaacH8sr
% ps0lbbf9q8WrFfeuY-ribbf9v8qqaqFr0xc9pk0xbba9q8WqFfea0-
% yr0RYxir-Jbba9q8aq0-yq-He9q8qqQ8frFve9Fve9Ff0dc9Gqpi0d
% meaabaqaciGacaGaaeqabaWaaeaaeaaakeaadaWcaaqaaiabgkGi2k
% abeY7aTnaaBaaaleaacaaIWaaabeaakiaadIeadaWgaaWcbaGaamOE
% aaqabaaakeaacqGHciITcaWG6baaaiabg2da9iabgkHiTmaalaaaba
% GaeyOaIyRaeqiVd02aaSbaaSqaaiaaicdaaeqaaOGaamytamaaBaaa
% leaacaWG6baabeaaaOqaaiabgkGi2kaadQhaaaGaeyypa0JaamOqam
% aaBaaaleaacaWG6baabeaakmaalaaabaGaaGymaaqaaiaaikdaaaWa
% aSaaaeaacqGHciITcqaHYoGyaeaacqGHciITcaWG6baaaaaa!4E82!
\frac{{\partial {\mu _0}{H_z}}}{{\partial z}} =  - \frac{{\partial {\mu _0}{M_z}}}{{\partial z}} = {B_z}\frac{1}{2}\frac{{\partial \beta }}{{\partial z}} \ \ ,
\end{equation}
where $\beta$ is computed with the electron density. If we assume that this density is equal to the neutral hydrogen density, following our proposal, we obtain  
% MathType!MTEF!2!1!+-
% faaahqart1ev3aaaKnaaaaWenf2ys9wBH5garuavP1wzZbItLDhis9
% wBH5garmWu51MyVXgaruWqVvNCPvMCaebbnrfifHhDYfgasaacH8sr
% ps0lbbf9q8WrFfeuY-ribbf9v8qqaqFr0xc9pk0xbba9q8WqFfea0-
% yr0RYxir-Jbba9q8aq0-yq-He9q8qqQ8frFve9Fve9Ff0dc9Gqpi0d
% meaabaqaciGacaGaaeqabaWaaeaaeaaakeaacaWGcbWaaSbaaSqaai
% aadQhaaeqaaOWaaSGbaeaacqGHciITcqaHYoGyaeaacaaIYaGaeyOa
% IyRaamOEaaaacqGHijYUcqGHsislcaaIZaGaaiOlaiaaisdaaaa!3E84!
${B_z}{{\partial \beta } \mathord{\left/
 {\vphantom {{\partial \beta } {2\partial z}}} \right.
 \kern-\nulldelimiterspace} {2\partial z}} \approx  - 3.4$
G/km for $B=1000$ G (and $-1.4$ G/km for $B=2500$ G) at the formation height of the \ion{Fe}{i} lines used for the measurements and for the neutral hydrogen density of the Maltby-M sunspot umbra model, which is comparable to the observed value 
% MathType!MTEF!2!1!+-
% faaahqart1ev3aaaKnaaaaWenf2ys9wBH5garuavP1wzZbItLDhis9
% wBH5garmWu51MyVXgaruWqVvNCPvMCaebbnrfifHhDYfgasaacH8sr
% ps0lbbf9q8WrFfeuY-ribbf9v8qqaqFr0xc9pk0xbba9q8WqFfea0-
% yr0RYxir-Jbba9q8aq0-yq-He9q8qqQ8frFve9Fve9Ff0dc9Gqpi0d
% meaabaqaciGacaGaaeqabaWaaeaaeaaakeaadaWcgaqaaiabgkGi2k
% abeY7aTnaaBaaaleaacaaIWaaabeaakiaadIeadaWgaaWcbaGaamOE
% aaqabaaakeaacqGHciITcaWG6baaaiabgIKi7kabgkHiTiaaiodaaa
% a!3D63!
${{\partial {\mu _0}{H_z}} \mathord{\left/
 {\vphantom {{\partial {\mu _0}{H_z}} {\partial z}}} \right.
 \kern-\nulldelimiterspace} {\partial z}} \approx  - 3$ 
G/km. This value is $-1.2$ G/km for the quiet Sun VALC model and for $B=1000$ G.

The case of laboratory plasmas is very different. From Eq. (\ref{eq--diamagnetism}), when $\vec{M}$ is comparable to $\vec{H}$ and $\vec{B}/\mu_0$, the plasma $\beta$ is on the order of unity (we temporarily assume here that  plasma is made of free charged particles). When $\beta$ is on the order of unity, the gas pressure is comparable to the magnetic pressure. Therefore, particles can escape under the effect of the gas pressure, and the magnetic field will be insufficient to keep then. $\vec{M}$ is an obstacle to plasma confinement \citep{Delcroix-Bers-94}. Consequently, in usually observed laboratory plasmas, $\vec{M}$ probably remains weak with respect to $\vec{H}$ and $\vec{B}/\mu_0$. Therefore,  we have $\vec{B} \approx \mu_0\vec{H}$;  $\vec{B}$ and $\mu_0\vec{H}$ are equivalent in these plasmas. In this paper we consider a plasma where $\vec{M}$, $\vec{H}$, and $\vec{B}/\mu_0$ are all comparable, and we study the effect of $\vec{M}$ on this plasma. This plasma is the solar photosphere and is naturally formed and maintained, but it is different from the laboratory confined plasmas.

\subsection{Magnetic field produced by the conduction currents}

The contribution $\vec{H}_c$ to the magnetic field due to the (macroscopic) conduction currents $\vec{J}$ is 
\begin{equation}
% MathType!MTEF!2!1!+-
% faaahqart1ev3aaaKnaaaaWenf2ys9wBH5garuavP1wzZbItLDhis9
% wBH5garmWu51MyVXgaruWqVvNCPvMCaebbnrfifHhDYfgasaacH8sr
% ps0lbbf9q8WrFfeuY-ribbf9v8qqaqFr0xc9pk0xbba9q8WqFfea0-
% yr0RYxir-Jbba9q8aq0-yq-He9q8qqQ8frFve9Fve9Ff0dc9Gqpi0d
% meaabaqaciGacaGaaeqabaWaaeaaeaaakeaacuGHhis0gaWcaiabgE
% na0kqadIeagaWcamaaBaaaleaacaWGJbaabeaakiabg2da9iqadQea
% gaWcaaaa!3912!
\vec \nabla  \times {\vec H_c} = \vec J \ \ ,
\end{equation}
such that $\vec{H}_c$ can be computed by applying the Biot \& Savart law to $\vec{J}$. As a consequence, $\vec{H}_c$ is a divergence-free field 
\begin{equation}
% MathType!MTEF!2!1!+-
% faaahqart1ev3aaaKnaaaaWenf2ys9wBH5garuavP1wzZbItLDhis9
% wBH5garmWu51MyVXgaruWqVvNCPvMCaebbnrfifHhDYfgasaacH8sr
% ps0lbbf9q8WrFfeuY-ribbf9v8qqaqFr0xc9pk0xbba9q8WqFfea0-
% yr0RYxir-Jbba9q8aq0-yq-He9q8qqQ8frFve9Fve9Ff0dc9Gqpi0d
% meaabaqaciGacaGaaeqabaWaaeaaeaaakeaacuGHhis0gaWcaiabgw
% SixlqadIeagaWcamaaBaaaleaacaWGJbaabeaakiabg2da9iaaicda
% aaa!391E!
\vec \nabla  \cdot {\vec H_c} = 0 \ \ .
\end{equation}

Generally,  however, we have in magnetized media
\begin{equation}
\vec{H} = \vec{H}_c + \vec{H}_d \ \ ,
\label{eq--total-field}
\end{equation}
where $\vec{H}_d$ is a curl-free field
\begin{equation}
% MathType!MTEF!2!1!+-
% faaahqart1ev3aaaKnaaaaWenf2ys9wBH5garuavP1wzZbItLDhis9
% wBH5garmWu51MyVXgaruWqVvNCPvMCaebbnrfifHhDYfgasaacH8sr
% ps0lbbf9q8WrFfeuY-ribbf9v8qqaqFr0xc9pk0xbba9q8WqFfea0-
% yr0RYxir-Jbba9q8aq0-yq-He9q8qqQ8frFve9Fve9Ff0dc9Gqpi0d
% meaabaqaciGacaGaaeqabaWaaeaaeaaakeaacuGHhis0gaWcaiabgE
% na0kqadIeagaWcamaaBaaaleaacaWGKbaabeaakiabg2da9iaaicda
% aaa!38EC!
\vec \nabla  \times {\vec H_d} = 0
,\end{equation}
as we introduce in the following section.

\subsection{Magnetic masses}
There is another contribution to the magnetic field that  is due to the magnetization resulting from the plasma diamagnetism. This magnetization $\vec{M}$ is a magnetic moment density, which is continuously spread outside the atom. In the solar photosphere and following our proposal of electron accumulation in the solar photosphere, we have
\begin{equation}
% MathType!MTEF!2!1!+-
% faaahqart1ev3aaaKnaaaaWenf2ys9wBH5garuavP1wzZbItLDhis9
% wBH5garmWu51MyVXgaruWqVvNCPvMCaebbnrfifHhDYfgasaacH8sr
% ps0lbbf9q8WrFfeuY-ribbf9v8qqaqFr0xc9pk0xbba9q8WqFfea0-
% yr0RYxir-Jbba9q8aq0-yq-He9q8qqQ8frFve9Fve9Ff0dc9Gqpi0d
% meaabaqaciGacaGaaeqabaWaaeaaeaaakeaacuGHhis0gaWcaiabgw
% Sixlqad2eagaWcaiabgcMi5kaaicdaaaa!38C6!
\vec \nabla  \cdot \vec M \ne 0
\end{equation}
because of the vertical density gradient. We can then introduce
\begin{equation}
% MathType!MTEF!2!1!+-
% faaahqart1ev3aaaKnaaaaWenf2ys9wBH5garuavP1wzZbItLDhis9
% wBH5garmWu51MyVXgaruWqVvNCPvMCaebbnrfifHhDYfgasaacH8sr
% ps0lbbf9q8WrFfeuY-ribbf9v8qqaqFr0xc9pk0xbba9q8WqFfea0-
% yr0RYxir-Jbba9q8aq0-yq-He9q8qqQ8frFve9Fve9Ff0dc9Gqpi0d
% meaabaqaciGacaGaaeqabaWaaeaaeaaakeaacuGHhis0gaWcaiabgw
% SixlqadIeagaWcamaaBaaaleaacaWGKbaabeaakiabg2da9iabgkHi
% TiqbgEGirBaalaGaeyyXICTabmytayaalaGaeyypa0JaeqyWdi3aaS
% baaSqaaiaad2eaaeqaaaaa!41DC!
\vec \nabla  \cdot {\vec H_d} =  - \vec \nabla  \cdot \vec M = {\rho _M} \ \ ,
\label{eq--rho0}
\end{equation}
where the curl-free field $\vec{H}_d$ has the mathematical form of an electric or gravitation field due to the magnetic mass density $\rho_M$, which may be positive or negative. The magnetic mass density is defined by the leftmost member of the above Eq. (\ref{eq--rho0}). We  recall that the magnetic mass is a mathematical tool only.

In the case of a finite magnetized volume, surface magnetic masses are formed on the volume surface. Denoting by $\vec{n}$ the unit vector perpendicular to the surface and oriented  outward, the surface magnetic mass density is
\begin{equation}
% MathType!MTEF!2!1!+-
% faaahqart1ev3aaaKnaaaaWenf2ys9wBH5garuavP1wzZbItLDhis9
% wBH5garmWu51MyVXgaruWqVvNCPvMCaebbnrfifHhDYfgasaacH8sr
% ps0lbbf9q8WrFfeuY-ribbf9v8qqaqFr0xc9pk0xbba9q8WqFfea0-
% yr0RYxir-Jbba9q8aq0-yq-He9q8qqQ8frFve9Fve9Ff0dc9Gqpi0d
% meaabaqaciGacaGaaeqabaWaaeaaeaaakeaacqaHdpWCdaWgaaWcba
% GaamytaiaadofaaeqaaOGaeyypa0JabmytayaalaGaeyyXICTabmOB
% ayaalaaaaa!3A5B!
{\sigma _{MS}} = \vec M \cdot \vec n \ \ .
\end{equation}

We consider again the oversimplified model of a purely vertical field or induction along $z$. We  assume that $B_z$ is positive and constant. The magnetization resulting from the plasma diamagnetism in Eq. (\ref{eq--diamagnetism}) is then also vertical, but with $M_z$ negative and with modulus decreasing with height. As a result $\rho_M > 0$. However, if $\rho_M$ is spatially constant, the resulting field $\vec{H}_d$ is zero, due to reasons of symmetry  and Eq. (\ref{eq--rho0}) cannot be satisfied. But this is not the case: the $\beta$ gradient itself decreases with height in such a way that $\rho_M > 0$ decreases with height. As a result $\vec{H}_d$ is oriented upward, and decreases with height. The total field $\vec{H}$ of Eq. (\ref{eq--total-field})  then decreases with height and has a non-zero divergence as observed, provided that it is proven that what is measured is the magnetic field $\vec{H}$ and not the magnetic induction $\vec{B}$, which is the object of the next section.

The $\vec{H}_d$ field is called the demagnetising field because it is opposite to $\vec{M}$.

\subsection{Contribution of the magnetization current}

A magnetization current appears when 
\begin{equation}
% MathType!MTEF!2!1!+-
% faaahqart1ev3aaaKnaaaaWenf2ys9wBH5garuavP1wzZbItLDhis9
% wBH5garmWu51MyVXgaruWqVvNCPvMCaebbnrfifHhDYfgasaacH8sr
% ps0lbbf9q8WrFfeuY-ribbf9v8qqaqFr0xc9pk0xbba9q8WqFfea0-
% yr0RYxir-Jbba9q8aq0-yq-He9q8qqQ8frFve9Fve9Ff0dc9Gqpi0d
% meaabaqaciGacaGaaeqabaWaaeaaeaaakeaacuGHhis0gaWcaiabgE
% na0kqad2eagaWcaiabg2da9iqadQeagaWcamaaBaaaleaacaWGnbaa
% beaakiabgcMi5kaaicdaaaa!3B82!
\vec \nabla  \times \vec M = {\vec J_M} \ne 0 \ \ ,
\end{equation}
which is the case when there are density variations such that the small loops (made by the charged particles about the magnetic induction and responsible for the magnetization) do not counterbalance. We then have  
\begin{equation}
% MathType!MTEF!2!1!+-
% faaahqart1ev3aaaKnaaaaWenf2ys9wBH5garuavP1wzZbItLDhis9
% wBH5garmWu51MyVXgaruWqVvNCPvMCaebbnrfifHhDYfgasaacH8sr
% ps0lbbf9q8WrFfeuY-ribbf9v8qqaqFr0xc9pk0xbba9q8WqFfea0-
% yr0RYxir-Jbba9q8aq0-yq-He9q8qqQ8frFve9Fve9Ff0dc9Gqpi0d
% meaabaqaciGacaGaaeqabaWaaeaaeaaakeaacuGHhis0gaWcaiabgE
% na0kqadkeagaWcaiabg2da9iabeY7aTnaaBaaaleaacaaIWaaabeaa
% kmaabmaabaGabmOsayaalaGaey4kaSIabmOsayaalaWaaSbaaSqaai
% aad2eaaeqaaaGccaGLOaGaayzkaaaaaa!3EE8!
\vec \nabla  \times \vec B = {\mu _0}\left( {\vec J + {{\vec J}_M}} \right) \ \ ,
\end{equation}
so that $\vec{B}$ can be computed by applying the Biot \& Savart law to $\vec{J}$ and $\vec{J}_M$.

In the case of a finite magnetized volume, surface magnetization currents are formed on the volume surface. Denoting  as $\vec{n}$ the unit vector perpendicular to the surface and oriented outward, the surface magnetization current density is
\begin{equation}
% MathType!MTEF!2!1!+-
% faaahqart1ev3aaaKnaaaaWenf2ys9wBH5garuavP1wzZbItLDhis9
% wBH5garmWu51MyVXgaruWqVvNCPvMCaebbnrfifHhDYfgasaacH8sr
% ps0lbbf9q8WrFfeuY-ribbf9v8qqaqFr0xc9pk0xbba9q8WqFfea0-
% yr0RYxir-Jbba9q8aq0-yq-He9q8qqQ8frFve9Fve9Ff0dc9Gqpi0d
% meaabaqaciGacaGaaeqabaWaaeaaeaaakeaaceWGkbGbaSaadaWgaa
% WcbaGaamytaiaadofaaeqaaOGaeyypa0JabmytayaalaGaey41aqRa
% bmOBayaalaaaaa!3946!
{\vec J_{MS}} = \vec M \times \vec n \ \ .
\end{equation}

The effect of $\vec{M}$ on $\vec{H}$ or $\vec{B}$ can be evaluated either via the magnetic masses and their demagnetising field, or via the Biot \& Savart law applied to the magnetization current. The two approaches lead to the same result, as  can be easily verified in the case of a sphere with constant magnetization (also considering  the surface magnetic masses and magnetization current). 

\section{The Zeeman Hamiltonian}
\label{sect--Zeeman}

\begin{figure}
\resizebox{\hsize}{!}{\includegraphics{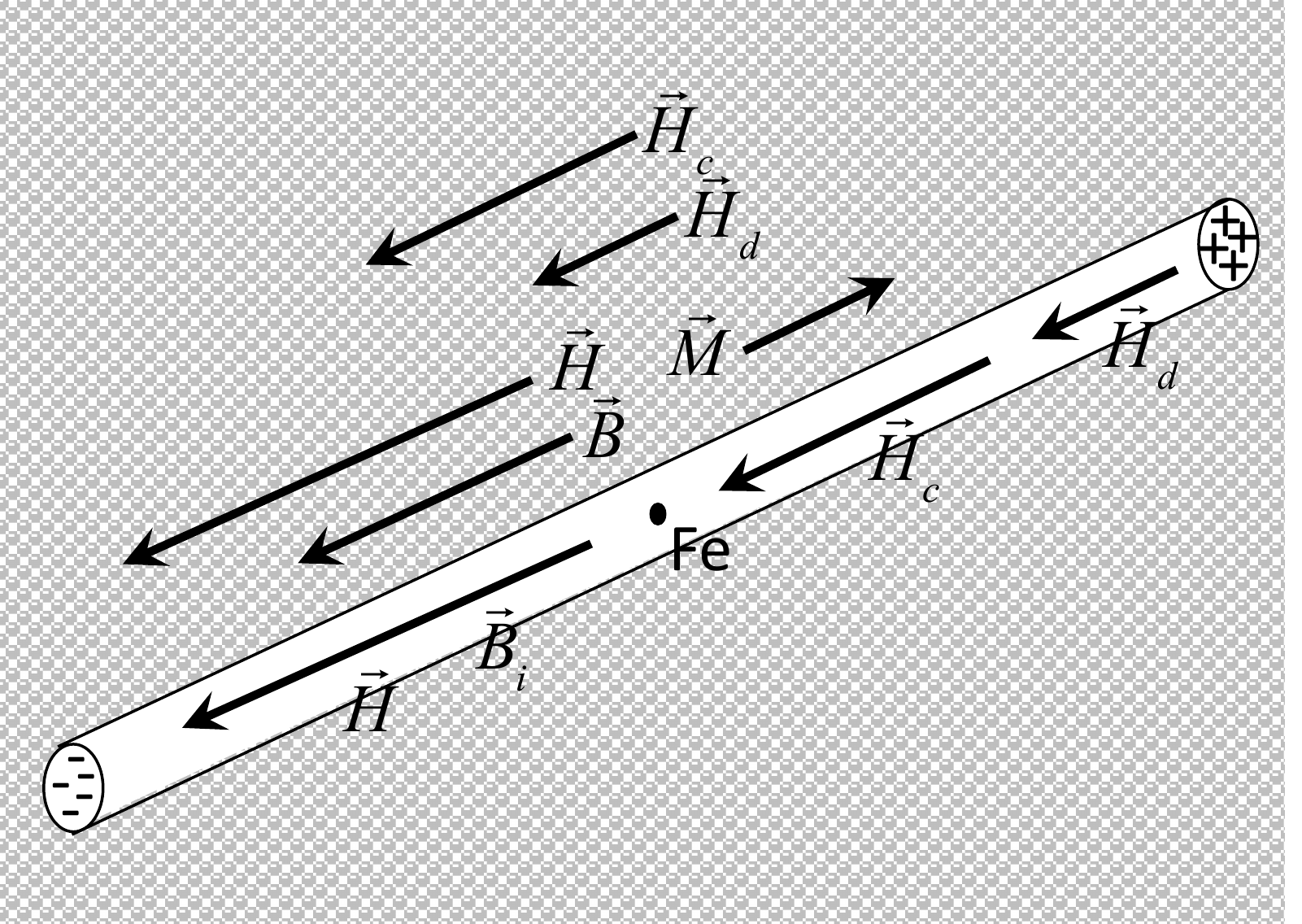}}
\caption{Microscopic local vacuum about the Fe atom, embedded in matter with magnetization $\vec{M}$. Because under strong magnetic conditions like in sunspot fields, matter goes along the magnetic field, the vacuum about the Fe atom has the form of a cylinder.}
\label{cylindre}
\end{figure}

The Zeeman Hamiltonian is the interaction energy between the atom having an elementary magnetic momentum $\vec{m}$ and the magnetic field.

In atomic physics, this Hamiltonian is obtained in particular by developing the atom impulse in the presence of a magnetic potential $(\vec{P}-q\vec{A})^2$, which leads to the well-known Zeeman Hamiltonian $-\vec{m} \cdot \vec {B}$, but the atom here considered lies in a vacuum. In the present paper we study the case of an atom embedded in a magnetized material, and we study in particular the effect of the surrounding material on the atom.

In the following, we present four demonstrations all agreeing that the Zeeman Hamiltonian for the atom embedded in the magnetized material is $-\mu_0 \vec{m} \cdot \vec {H}$, where the magnetic field $\vec{H}$ includes the demagnetising field $\vec{H}_d$ created by the surrounding magnetization, whose effect on the atom is thus taken into account. This is a result currently used in magnetized materials physics  \citep[see, e.g.,][]{Gignoux-Schmitt-93,Garnier-etal-98,Zhang-etal-94}.

\subsection{Demonstrations 1 and 2: evacuating an elementary matter cylinder to determine its magnetic energy}

The aim of this demonstration is to determine the energy required to evacuate to infinity or, alternatively, to fill from infinity a cylinder of matter containing  a single atom of magnetic momentum $\vec{m}$. The form of the elementary volume is determined by the vector direction of the field to which the atom is submitted in the matter. This demonstration is inspired by \citet{Gignoux-etal-02}, Section 2.2.2 of Chapter 2.

At the microscopic scale, the medium is made of particles (free electrons, ions, and neutral atoms, themselves made of the nucleus and electrons), and the Maxwell law Eq. (\ref{eq--BHM}) is the result of a macroscopic averaging as described in \citet{Jackson-75}, Sect. 6.7, ``Derivation of the Equations of Macroscopic Electromagnetism.'' This averaging is performed on lengths that are large with respect to the microscopic scales. As a result the magnetization is represented by the quantity $\vec{M}$, which is a density of magnetic moments ({i.e.,} a spatially smoothed quantity in cm$^{-3}$; see Eq. (\ref{M-n}) below), whereas the medium is made of particles at the microscopic scale.

In order to determine at what field the atom is submitted in the medium, it is then necessary to go back from the macroscopic smoothed scale to the microscopic scale where the atom is in a vacuum, with the other atoms or ions or free electrons far from it. This atom is however submitted to the field created by the other particles even far from it.

Considering that in sunspot umbra the charged particles move along the magnetic field and drag  the neutral particles along with them in such a way that there is no displacement perpendicular to the field in a first approximation, the vacuum created around the atom takes the form of a very long cylinder along the magnetic field, as represented in Fig. \ref{cylindre}. In this figure, we have also represented the magnetic field $\vec{H}_c$ due to the conduction currents (see previous section), and the demagnetising magnetic field $\vec{H}_d$ due to the non-zero $\mathrm{div}\vec{M}$, which is the way the magnetization affects the atom.

The magnetic field variations in induction units increases up to 3 G/km following the observations, which corresponds to a variation of $3 \times 10^{-10}$ G along a scale of 1000 \AA. This variation can be neglected at the cylinder diameter scale, so that $\vec{B}$, $\vec{H}$, and $\vec{M}$ can be considered   constant close to the atom and to the cylinder center.

The cylinder dimensions are assumed to be large with respect to the atom dimensions, but small with respect to the macroscopic characteristic lengths, and such that the cylinder contains only one atom to be evacuated to infinity in order to determine its magnetic energy. The cylinder length is assumed to be very large with respect to the cylinder width. Magnetic masses appear at the cylinder extremities where the separation surface is not parallel to $\vec{M}$. However, the second demagnetising field created by these masses is negligible at the cylinder center because the cylinder is very long. This may be the second reason for considering the cylinder form to describe the vacuum volume made around the atom. For any other form, for instance a sphere, such magnetic masses would also be present and would induce a non-negligible second demagnetising field, which would also be present outside the volume and would add   a perturbation to the local field, induction, and magnetization. The long cylinder form (or at least an elongated form parallel to the field) is the only possible form  considered that does not add any perturbation to the local fields. A third argument in favor of the cylinder form is the symmetry of the medium, which is cylindrical along the field, induction, and magnetization, therefore not spherical.

\subsubsection{Demonstration 1: via magnetic masses}

Outside the cylinder, the magnetic induction is $\vec{B}$, the magnetization is $\vec{M}$ and the magnetic field is $\vec{H}=\vec{H}_c+\vec{H}_d$. Inside the cylinder, close to its center, the tangential component of $\vec{H}$ and the normal component of $\vec{B}$ are transmitted through the separation surface. The cylinder axis is parallel to $\vec{B}$ and to $\vec{H}$, so that $\vec{H}$ is transmitted inside the cylinder. There is no matter inside the cylinder, which is assumed to be empty for single atom energy study purposes; therefore, no magnetization is present. Therefore, the magnetic induction inside the cylinder is
\begin{equation}
\vec{B}_i = \mu_0 \vec{H} \ \,
\label{eq--B_i}
,\end{equation}
which is different from the induction $\vec{B}$ outside the cylinder, where there is some magnetization and where $\vec{B}=\mu_0\left(\vec{H}+\vec{M}\right)$. We  assume constant $\vec{H}$  at the atom scale, so  at this scale, which is the one of our present problem, $\mathrm{div}\vec{B}_i = 0$ is satisfied. 

The cylinder is parallel to $\vec{B}$ and to $\vec{H}$. As a consequence, the free electrons, which move along the magnetic field, do not enter the cylinder, which contains a single atom. We note, however, that the electron Larmor radius in typical photospheric fields is about 25 microns, which is large with respect to the atomic or microscopic typical dimensions. However, the magnetization due to these electrons was averaged as $\vec{M}$, which affects the atom via $\vec{H}_d$, which is transmitted inside the cylinder.

\subsubsection{Demonstration 2: via magnetization currents}

As described in Sect. 1.2.3 of Chapter 2 of \citet{Gignoux-etal-02}, the magnetic induction $\vec{B}_i$ inside the empty cylinder can also be evaluated by considering the magnetic induction $\Delta \vec{B}$ created by the matter, which is evacuated 
in order to create the vacuum in the cylinder. The aim of this operation is   to determine the matter cylinder energy in the field by bringing back the matter cylinder from infinity to the empty cylinder. This matter cylinder is made of constant magnetization $\vec{M}$, and the magnetic induction inside a cylinder of constant magnetization $\vec{M}$ is $\Delta \vec{B} = \mu_0 \vec{M}$ \citep[see, e.g.,][Sect. 1.1.6]{Gignoux-etal-02}. This magnetic induction can be computed from the magnetization currents at the surface of the cylinder. Therefore inside the vacuum cylinder, the induction is $\vec{B}_i = \vec{B} - \Delta \vec{B}$, and Eq. (\ref{eq--B_i}) is recovered.

\subsubsection{Agreement of the two demonstrations}

These two determinations of $\vec{B}_i$, one based on the magnetic masses evaluation, and the second one based on the magnetization currents evaluation, fully agree, as expected.

The energy of the atomic magnetic momentum denoted as $\vec{m}$ is then
\begin{equation}
% MathType!MTEF!2!1!+-
% faaahqart1ev3aaaKnaaaaWenf2ys9wBH5garuavP1wzZbItLDhis9
% wBH5garmWu51MyVXgaruWqVvNCPvMCaebbnrfifHhDYfgasaacH8sr
% ps0lbbf9q8WrFfeuY-ribbf9v8qqaqFr0xc9pk0xbba9q8WqFfea0-
% yr0RYxir-Jbba9q8aq0-yq-He9q8qqQ8frFve9Fve9Ff0dc9Gqpi0d
% meaabaqaciGacaGaaeqabaWaaeaaeaaakeaacaWGxbGaeyypa0Jaey
% OeI0IabmyBayaalaGaeyyXICTabmOqayaalaWaaSbaaSqaaiaadMga
% aeqaaOGaeyypa0JaeyOeI0IaeqiVd02aaSbaaSqaaiaaicdaaeqaaO
% GabmyBayaalaGaeyyXICTabmisayaalaaaaa!425F!
W =  - \vec m \cdot {\vec B_i} =  - {\mu _0}\vec m \cdot \vec H \ \ .
\end{equation}

\subsection{Demonstration 3: from the magnetic potential energy}

The total magnetic energy of a medium with magnetic field $\vec{H}$ and induction $\vec{B}$, which is either paramagnetic or diamagnetic, as ours is, can be written \citep[][Sect. 6.2, ``Energy in the Magnetic Field'']{Jackson-75}
\begin{equation}
% MathType!MTEF!2!1!+-
% faaahqart1ev3aaaKnaaaaWenf2ys9wBH5garuavP1wzZbItLDhis9
% wBH5garmWu51MyVXgaruWqVvNCPvMCaebbnrfifHhDYfgasaacH8sr
% ps0lbbf9q8WrFfeuY-ribbf9v8qqaqFr0xc9pk0xbba9q8WqFfea0-
% yr0RYxir-Jbba9q8aq0-yq-He9q8qqQ8frFve9Fve9Ff0dc9Gqpi0d
% meaabaqaciGacaGaaeqabaWaaeaaeaaakeaacaWGxbGaeyypa0Zaa8
% qaaeaadaWcaaqaaiqadIeagaWcaiabgwSixlqadkeagaWcaaqaaiaa
% ikdaaaGaciizamaaCaaaleqabaGaaG4maaaakiqadkhagaWcaaWcbe
% qab0Gaey4kIipaaaa!3D11!
W = \int {\frac{{\vec H \cdot \vec B}}{2}{{\mathop{\rm d}\nolimits} ^3}\vec r} \ \ .
\end{equation}

By applying Eq. (\ref{eq--BHM}), this integral can be expanded as
\begin{equation}
% MathType!MTEF!2!1!+-
% faaahqart1ev3aaaKnaaaaWenf2ys9wBH5garuavP1wzZbItLDhis9
% wBH5garmWu51MyVXgaruWqVvNCPvMCaebbnrfifHhDYfgasaacH8sr
% ps0lbbf9q8WrFfeuY-ribbf9v8qqaqFr0xc9pk0xbba9q8WqFfea0-
% yr0RYxir-Jbba9q8aq0-yq-He9q8qqQ8frFve9Fve9Ff0dc9Gqpi0d
% meaabaqaciGacaGaaeqabaWaaeaaeaaakeaacaWGxbGaeyypa0Zaa8
% qaaeaadaWcaaqaaiabeY7aTnaaBaaaleaacaaIWaaabeaakiaadIea
% daahaaWcbeqaaiaaikdaaaaakeaacaaIYaaaaiGacsgadaahaaWcbe
% qaaiaaiodaaaGcceWGYbGbaSaaaSqabeqaniabgUIiYdGccqGHRaWk
% daWdbaqaamaalaaabaGaeqiVd02aaSbaaSqaaiaaicdaaeqaaOGabm
% ytayaalaGaeyyXICTabmisayaalaaabaGaaGOmaaaaciGGKbWaaWba
% aSqabeaacaaIZaaaaOGabmOCayaalaaaleqabeqdcqGHRiI8aaaa!4AC2!
W = \int {\frac{{{\mu _0}{H^2}}}{2}{{\mathop{\rm d}\nolimits} ^3}\vec r}  + \int {\frac{{{\mu _0}\vec M \cdot \vec H}}{2}{{\mathop{\rm d}\nolimits} ^3}\vec r} \ \ ,
\end{equation}
where the first  member on the right-hand side is the field energy and the second   is the magnetization, or magnetized matter, energy in the field.

The magnetization $\vec{M}$ is   the spatial average of local elementary magnetic moments \citep[][Eq. (6.98)]{Jackson-75}
\begin{equation}
% MathType!MTEF!2!1!+-
% faaahqart1ev3aaaKnaaaaWenf2ys9wBH5garuavP1wzZbItLDhis9
% wBH5garmWu51MyVXgaruWqVvNCPvMCaebbnrfifHhDYfgasaacH8sr
% ps0lbbf9q8WrFfeuY-ribbf9v8qqaqFr0xc9pk0xbba9q8WqFfea0-
% yr0RYxir-Jbba9q8aq0-yq-He9q8qqQ8frFve9Fve9Ff0dc9Gqpi0d
% meaabaqaciGacaGaaeqabaWaaeaaeaaakeaaceWGnbGbaSaadaqada
% qaaiqadkhagaWcaaGaayjkaiaawMcaaiabg2da9maaamaabaWaaabu
% aeaaceWGTbGbaSaadaWgaaWcbaGaamOBaaqabaGccqaH0oazdaqada
% qaaiqadkhagaWcaiabgkHiTiqadkhagaWcamaaBaaaleaacaWGUbaa
% beaaaOGaayjkaiaawMcaaaWcbaGaamOBaaqab0GaeyyeIuoaaOGaay
% zkJiaawQYiaaaa!446E!
\vec M\left( {\vec r} \right) = \left\langle {\sum\limits_n {{{\vec m}_n}\delta \left( {\vec r - {{\vec r}_n}} \right)} } \right\rangle \ \ ,
\label{M-n}
\end{equation}
where $\delta$ is the Dirac function and where $\left\langle \right\rangle$ is the spatial average.

By considering an elementary volume about the atom of magnetic moment $\vec{m}$, with only the atom inside the volume, the atom energy is then, in terms of the total energy,
\begin{equation}
% MathType!MTEF!2!1!+-
% faaahqart1ev3aaaKnaaaaWenf2ys9wBH5garuavP1wzZbItLDhis9
% wBH5garmWu51MyVXgaruWqVvNCPvMCaebbnrfifHhDYfgasaacH8sr
% ps0lbbf9q8WrFfeuY-ribbf9v8qqaqFr0xc9pk0xbba9q8WqFfea0-
% yr0RYxir-Jbba9q8aq0-yq-He9q8qqQ8frFve9Fve9Ff0dc9Gqpi0d
% meaabaqaciGacaGaaeqabaWaaeaaeaaakeaacqaH0oazcaWGxbGaey
% ypa0ZaaSaaaeaacqaH8oqBdaWgaaWcbaGaaGimaaqabaGcceWGTbGb
% aSaacqGHflY1ceWGibGbaSaaaeaacaaIYaaaaaaa!3CA5!
\delta W = \frac{{{\mu _0}\vec m \cdot \vec H}}{2} \ \ .
\end{equation}
However, the system is not totally isolated because there are external current sources. When a small matter piece is introduced, or a small displacement is performed, energy is transferred from the sources to the system. In his Sect. 6.2, \citet{Jackson-75} writes that ``we can show that for a small displacement the work done against the induced electromotive forces is twice as large as, and of the opposite sign to, the potential-energy change of the body.'' The energy of the atom embedded in the magnetized matter is therefore
\begin{equation}
% MathType!MTEF!2!1!+-
% faaahqart1ev3aaaKnaaaaWenf2ys9wBH5garuavP1wzZbItLDhis9
% wBH5garmWu51MyVXgaruWqVvNCPvMCaebbnrfifHhDYfgasaacH8sr
% ps0lbbf9q8WrFfeuY-ribbf9v8qqaqFr0xc9pk0xbba9q8WqFfea0-
% yr0RYxir-Jbba9q8aq0-yq-He9q8qqQ8frFve9Fve9Ff0dc9Gqpi0d
% meaabaqaciGacaGaaeqabaWaaeaaeaaakeaacqaH0oazcaWGxbGaey
% ypa0JaeyOeI0IaeqiVd02aaSbaaSqaaiaaicdaaeqaaOGabmyBayaa
% laGaeyyXICTabmisayaalaaaaa!3CC6!
\delta W =  - {\mu _0}\vec m \cdot \vec H \ \ ,
\end{equation}
where the effect of the surrounding dipoles on the atom is taken into account in the demagnetising field contribution $\vec{H}_d$ included in $\vec{H}$.

\subsection{Concluding remarks about the Zeeman Hamiltonian}

All the methods lead  to that conclusion that the energy of the atom embedded in the magnetized matter is $W =  - \mu _0 \vec{m} \cdot \vec{H}$. Accordingly, the Zeeman Hamiltonian $H_M$ describing the interaction between the atom (paramagnetic of momentum $\vec{m}$) and the magnetic field is
\begin{equation}
% MathType!MTEF!2!1!+-
% faaahqart1ev3aaaKnaaaaWenf2ys9wBH5garuavP1wzZbItLDhis9
% wBH5garmWu51MyVXgaruWqVvNCPvMCaebbnrfifHhDYfgasaacH8sr
% ps0lbbf9q8WrFfeuY-ribbf9v8qqaqFr0xc9pk0xbba9q8WqFfea0-
% yr0RYxir-Jbba9q8aq0-yq-He9q8qqQ8frFve9Fve9Ff0dc9Gqpi0d
% meaabaqaciGacaGaaeqabaWaaeaaeaaakeaacaWGibWaaSbaaSqaai
% aad2eaaeqaaOGaeyypa0JaeyOeI0IaeqiVd02aaSbaaSqaaiaaicda
% aeqaaOGabmyBayaalaGaeyyXICTabmisayaalaaaaa!3C1A!
{H_M} =  - {\mu _0}\vec m \cdot \vec H \ \ .
\end{equation}
When $LS$-coupling is valid for describing the atomic states, this may be simply rewritten as
\begin{equation}
% MathType!MTEF!2!1!+-
% faaahqart1ev3aaaKnaaaaWenf2ys9wBH5garuavP1wzZbItLDhis9
% wBH5garmWu51MyVXgaruWqVvNCPvMCaebbnrfifHhDYfgasaacH8sr
% ps0lbbf9q8WrFfeuY-ribbf9v8qqaqFr0xc9pk0xbba9q8WqFfea0-
% yr0RYxir-Jbba9q8aq0-yq-He9q8qqQ8frFve9Fve9Ff0dc9Gqpi0d
% meaabaqaciGacaGaaeqabaWaaeaaeaaakeaacaWGibWaaSbaaSqaai
% aad2eaaeqaaOGaeyypa0JaeyOeI0Iaam4zamaaBaaaleaacaWGkbaa
% beaakiabeY7aTnaaBaaaleaacaWGcbaabeaakiabeY7aTnaaBaaale
% aacaaIWaaabeaakiqadQeagaWcaiabgwSixlqadIeagaWcaaaa!409B!
{H_M} =  - {g_J}{\mu _B}{\mu _0}\vec J \cdot \vec H \ \ ,
\end{equation}
where $\vec{J}$ is the atomic total kinetic momentum $\vec{J} = \vec{L} + \vec{S}$, $g_J$ the level Land\'e factor and $\mu_B$ the Bohr magneton
% MathType!MTEF!2!1!+-
% faaahqart1ev3aaaKnaaaaWenf2ys9wBH5garuavP1wzZbItLDhis9
% wBH5garmWu51MyVXgaruWqVvNCPvMCaebbnrfifHhDYfgasaacH8sr
% ps0lbbf9q8WrFfeuY-ribbf9v8qqaqFr0xc9pk0xbba9q8WqFfea0-
% yr0RYxir-Jbba9q8aq0-yq-He9q8qqQ8frFve9Fve9Ff0dc9Gqpi0d
% meaabaqaciGacaGaaeqabaWaaeaaeaaakeaacqaH8oqBdaWgaaWcba
% GaamOqaaqabaGccqGH9aqpdaWcgaqaaiaadghacqWIpecAaeaacaaI
% YaGaamyBamaaBaaaleaacaWGLbaabeaaaaaaaa!3A31!
${\mu _B} = {{q\hbar } \mathord{\left/
 {\vphantom {{q\hbar } {2{m_e}}}} \right.
 \kern-\nulldelimiterspace} {2{m_e}}}$,
where $q$ and $m_e$ are the electron charge and mass, respectively.

As a result, the magnetic field $\vec{H}$, which includes the demagnetising field $\vec{H}_d$ contribution, governs the Zeeman Hamiltonian describing the interaction between the atom embedded in magnetized matter, and the magnetic field. This can be read, including the contribution of the demagnetising field, in \citet{Gignoux-Schmitt-93}, \citet{Garnier-etal-98}, and \citet{Zhang-etal-94}. The result is that what is measured by interpretation of the Zeeman effect in spectral lines in a magnetized medium like the solar photosphere, is $\vec{H}$, whose divergence may be non-zero at the macroscopic scale, as observed. 

Otherwise, it can be argued that $-\vec{m} \cdot \vec {B}$ is inappropriate to express the energy of the magnetic momentum $\vec{m}$ embedded in the magnetized matter and the magnetic field. When expanded following Eq. (\ref{eq--BHM}), products of the type $-\vec{m} \cdot \vec{M}$  appear, which  express, following Eq. (\ref{M-n}), that the matter acts on itself, even if the different values of  $\vec{m}$ are associated with different particles of it. As remarked by \citet[][Sect. 35]{Landau-Lifshitz-84}, when examining the case of a conductor carrying a current and placed in a magnetic field, ``the field of the conductor itself cannot, by the law of conservation of momentum, contribute to the total force acting on the conductor.'' This does not mean that the effect of the magnetic moments surrounding the atom is not taken into account; as previously stated, their effect is at the origin of the demagnetising field $\vec{H}_d$ included in the magnetic field $\vec{H}$, which enters the energy instead of $\vec{B}$.

\subsection{Demonstration 4: considering the mean atom submitted to the macroscopic mean magnetic field}

From the atomic point of view, the Zeeman Hamiltonian depends on the magnetic field external to the atom, which is the magnetic field at the atom position, with the exclusion of the possible field generated by the atom itself (the self-energy term must be excluded). Let us consider the mean atom submitted to the macroscopic mean field (see the averaging process described at the beginning of Sect. \ref{sect--fields}). As we consider the mean atom, $\vec{M}$ is made of the repeated atomic magnetic dipole, whose contribution to the Zeeman Hamiltonian should then   be ignored as an internal atomic contribution. As $\vec{M}$ is part of the magnetic induction $\vec{B}$ (see Eq. (\ref{eq--BHM})), this indicates that the Zeeman Hamiltonian is not governed by the magnetic induction $\vec{B}$, but by the magnetic field $\vec{H}$. The magnetic field generated by the surrounding dipoles is the demagnetising field $\vec{H}_d$ included in $\vec{H}$; the effect of the surrounding dipoles is therefore fully taken into account.

\citet{Langevin-05} demonstrated that the magnetic energy of a moment $\vec{M}$ placed in an external magnetic field $\vec{H}$ is $W = -\mu_0 \vec{M} \cdot \vec{H}$. Although he determined $\mathrm{div}\vec{H}=0$ for the external field in the exterior where $\vec{B} = \mu_0 \vec{H}$, the analysis of his experiment is unambiguous about the actual roles of $\vec{M}$, $\vec{B}$, and $\vec{H}$. In the case of non-negligible $\vec{M}$, $\vec{M}$ is contained in a magnetized barrel introduced in a coil  such that it is $\vec{H}$ and not $\vec{B}$ that is transmitted from the exterior to the interior of the barrel, because $\vec{M}$, $\vec{B}$, and $\vec{H}$ are all parallel to the barrel surface inside the coil, similarly to Fig. \ref{cylindre}. \citet{Weiss-07} showed that the effect of the surrounding dipoles is the demagnetising field, which contributes to $\vec{H}$ and is then fully taken into account in the magnetic energy.

\section{Electron accumulation at the surface of the Sun}
\label{sect--trap}

\begin{figure}
\resizebox{\hsize}{!}{\includegraphics{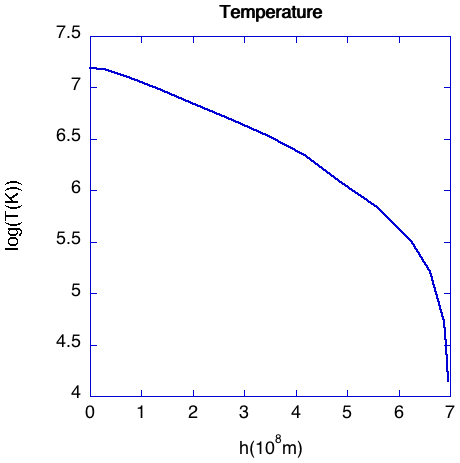}}
\resizebox{\hsize}{!}{\includegraphics{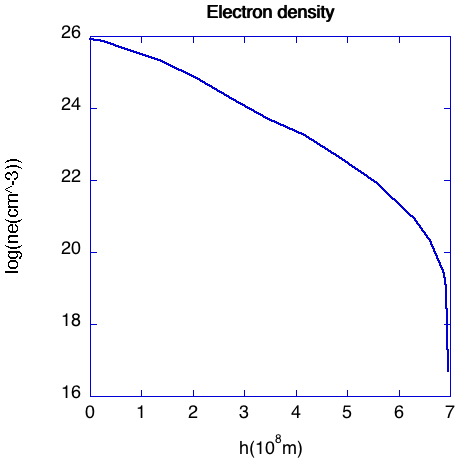}}
\caption{Temperature (top) and electron density (bottom) inside the Sun, from \citet{Allen-73} and assuming local electric neutrality. These figures show the different behaviors between the interior and the surface layer.}
\label{temp-dens}
\end{figure}

Our proposal is based on the consideration that electrons, which are much lighter than protons, have a thermal velocity inside the Sun much higher than the escape velocity, when protons remain submitted to the Sun's gravity. As a result, internal electrons tend to escape. The cloud of free electrons spreads out toward the star exterior. However, as we explain below, the escaping electrons come to be finally partly retained by the protons, the protons themselves are retained by the star gravity. The combined result is an accumulation of electrons in the star's more external layers, in particular the surface layer. It should be noted that the physical conditions in terms of temperature and density are specific in the surface layer (see  Fig. \ref{temp-dens}). We present and discuss this mechanism in the following.

\subsection{Escape velocity from gravity}

The fact that the electron thermal velocity surpasses the escape velocity is well known in the Solar Corona. As \citet{Meyer-Vernet-07} writes on p. 251, ``At a temperature of $10^6$ K, their thermal speed 
% MathType!MTEF!2!1!+-
% faaahqart1ev3aaaKnaaaaWenf2ys9wBH5garuavP1wzZbItLDhis9
% wBH5garmWu51MyVXgaruWqVvNCPvMCaebbnrfifHhDYfgasaacH8sr
% ps0lbbf9q8WrFfeuY-ribbf9v8qqaqFr0xc9pk0xbba9q8WqFfea0-
% yr0RYxir-Jbba9q8aq0-yq-He9q8qqQ8frFve9Fve9Ff0dc9Gqpi0d
% meaabaqaciGacaGaaeqabaWaaeaaeaaakeaacqWIdjYocaaI1aGaai
% OlaiaaiwdacqGHxdaTcaaIXaGaaGimamaaCaaaleqabaGaaGOnaaaa
% aaa!3959!
$ \simeq 5.5 \times {10^6}$
m/s is nearly 10 times greater than the escape speed.'' Instead, protons do not escape because ``electrons and protons have opposite charges, but their masses differ by the factor 
% MathType!MTEF!2!1!+-
% faaahqart1ev3aaaKnaaaaWenf2ys9wBH5garuavP1wzZbItLDhis9
% wBH5garmWu51MyVXgaruWqVvNCPvMCaebbnrfifHhDYfgasaacH8sr
% ps0lbbf9q8WrFfeuY-ribbf9v8qqaqFr0xc9pk0xbba9q8WqFfea0-
% yr0RYxir-Jbba9q8aq0-yq-He9q8qqQ8frFve9Fve9Ff0dc9Gqpi0d
% meaabaqaciGacaGaaeqabaWaaeaaeaaakeaadaWcgaqaaiaad2gada
% WgaaWcbaGaamiCaaqabaaakeaacaWGTbWaaSbaaSqaaiaadwgaaeqa
% aaaakiabloKi7iaaigdacaaI4aGaaG4maiaaiEdaaaa!39F0!
${{{m_p}} \mathord{\left/
 {\vphantom {{{m_p}} {{m_e}}}} \right.
 \kern-\nulldelimiterspace} {{m_e}}} \simeq 1837$,
so that electrons have a thermal speed greater than protons by a factor of order of magnitude 
% MathType!MTEF!2!1!+-
% faaahqart1ev3aaaKnaaaaWenf2ys9wBH5garuavP1wzZbItLDhis9
% wBH5garmWu51MyVXgaruWqVvNCPvMCaebbnrfifHhDYfgasaacH8sr
% ps0lbbf9q8WrFfeuY-ribbf9v8qqaqFr0xc9pk0xbba9q8WqFfea0-
% yr0RYxir-Jbba9q8aq0-yq-He9q8qqQ8frFve9Fve9Ff0dc9Gqpi0d
% meaabaqaciGacaGaaeqabaWaaeaaeaaakeaadaGcaaqaamaalyaaba
% GaamyBamaaBaaaleaacaWGWbaabeaaaOqaaiaad2gadaWgaaWcbaGa
% amyzaaqabaaaaaqabaGccqWIdjYocaaI0aGaaG4maaaa!3880!
$\sqrt {{{{m_p}} \mathord{\left/
 {\vphantom {{{m_p}} {{m_e}}}} \right.
 \kern-\nulldelimiterspace} {{m_e}}}}  \simeq 43$  
(because their temperatures have generally the same order of magnitude).''

Inside the Sun, at 0.5 $R_{\sun}$, the temperature is analogous, being $3.4 \times 10^6$ K, and 94\% of the solar mass is contained inside the sphere of 0.5 $R_{\sun}$ radius \citep{Allen-73}. If we recall that the escape velocity from a star is given by
\begin{equation}
% MathType!MTEF!2!1!+-
% faaahqart1ev3aaaKnaaaaWenf2ys9wBH5garuavP1wzZbItLDhis9
% wBH5garmWu51MyVXgaruWqVvNCPvMCaebbnrfifHhDYfgasaacH8sr
% ps0lbbf9q8WrFfeuY-ribbf9v8qqaqFr0xc9pk0xbba9q8WqFfea0-
% yr0RYxir-Jbba9q8aq0-yq-He9q8qqQ8frFve9Fve9Ff0dc9Gqpi0d
% meaabaqaciGacaGaaeqabaWaaeaaeaaakeaacaWG2bWaaSbaaSqaai
% aabwgacaqGZbGaae4yaaqabaGccqGH9aqpdaGcaaqaamaalaaabaGa
% aGOmaiaadEeacaWGnbWaaSbaaSqaaiabgYda8aqabaaakeaacaWGsb
% aaaaWcbeaaaaa!3B10!
{v_{{\rm{esc}}}} = \sqrt {\frac{{2G{M_ < }}}{R}} \ \ ,
\label{eq--gravity}
\end{equation}
where $G$ is the gravitation constant, $M_ <$ is the mass inside the sphere of radius $R$, where $R$ is the distance from the star's center where the escape velocity is evaluated. In these conditions, the electron thermal velocity is found to be 14 times higher than the escape velocity at 0.5 $R_{\sun}$, when the proton velocity is only 0.34 times  the escape velocity.

\subsection{Thermal escape}

Thermal escape is a well-known process in planet atmosphere studies \citep[see, e.g.,][for a review]{Chamberlain-63}. It is a very efficient mechanism for evaporating light elements like hydrogen from the Earth atmosphere, but the conditions in our case are very different  for two aspects. First, in these atmospheres the escape velocity remains significantly higher than the quadratic average thermal velocity defined as
\begin{equation}
% MathType!MTEF!2!1!+-
% faaahqart1ev3aaaKnaaaaWenf2ys9wBH5garuavP1wzZbItLDhis9
% wBH5garmWu51MyVXgaruWqVvNCPvMCaebbnrfifHhDYfgasaacH8sr
% ps0lbbf9q8WrFfeuY-ribbf9v8qqaqFr0xc9pk0xbba9q8WqFfea0-
% yr0RYxir-Jbba9q8aq0-yq-He9q8qqQ8frFve9Fve9Ff0dc9Gqpi0d
% meaabaqaciGacaGaaeqabaWaaeaaeaaakeaadaWcaaqaaiaaigdaae
% aacaaIYaaaaiaad2gacaWG2bWaa0baaSqaaiaabshacaqGObaabaGa
% aGOmaaaakiabg2da9maalaaabaGaaG4maaqaaiaaikdaaaGaam4Aam
% aaBaaaleaacaqGcbaabeaakiaadsfaaaa!3D1B!
\frac{1}{2}mv_{{\rm{th}}}^2 = \frac{3}{2}{k_{\rm{B}}}T \ \ ,
\end{equation}
which results in
\begin{equation}
% MathType!MTEF!2!1!+-
% faaahqart1ev3aaaKnaaaaWenf2ys9wBH5garuavP1wzZbItLDhis9
% wBH5garmWu51MyVXgaruWqVvNCPvMCaebbnrfifHhDYfgasaacH8sr
% ps0lbbf9q8WrFfeuY-ribbf9v8qqaqFr0xc9pk0xbba9q8WqFfea0-
% yr0RYxir-Jbba9q8aq0-yq-He9q8qqQ8frFve9Fve9Ff0dc9Gqpi0d
% meaabaqaciGacaGaaeqabaWaaeaaeaaakeaacaWG2bWaaSbaaSqaai
% aabshacaqGObaabeaakiabg2da9maakaaabaWaaSaaaeaacaaIZaGa
% am4AamaaBaaaleaacaqGcbaabeaakiaadsfaaeaacaWGTbaaaaWcbe
% aaaaa!3A36!
{v_{{\rm{th}}}} = \sqrt {\frac{{3{k_{\rm{B}}}T}}{m}} \ \ ,
\end{equation}
where $m$ is the mass of the particle under study (electron, proton), $k_{\rm{B}}$ is the Boltzmann constant, and $T$ the temperature. Then, in the case of planet atmospheres, only the tail of the velocity Maxwell distribution is affected by escaping. Second, escaping takes place in very low density regions, where the collision mean free path is larger than a typical length, such as the pressure scale height, which means that this atmosphere layer can be considered collision free. In the lower lying neighbor layer, however,  the few collisions  reestablish the thermal equilibrium and repopulate the escaping velocities. This is the source of the efficiency of the mechanism in planet atmospheres.

Our case of the solar interior is different   for these two points. First, the escape velocity is  much lower  than the electron quadratic average thermal velocity, by a factor of 14 at 0.5 $R_{\sun}$. As a consequence, nearly all the electron velocity Maxwell distribution is affected by escaping. Second, however, the 
mean free path for the electron-proton collision is about $7 \times 10^{-9}$ m at 0.5 $R_{\sun}$, when the density scale height is on the order of 55,000 km in the solar interior,  according to the  \citet{Allen-73} data. The electron-electron collision mean free path is about twice the electron-proton collision mean free path, as explained by \citet{Beck-MeyerV-08} in their note 5: ``For collisions between electrons the calculation must be done in the center-of-mass frame, and the mean free path for change in the velocity direction is greater by a factor of about 2.'' As the proton and electron densities are similar in the solar interior, the proton-proton collision mean free path is similar to that of the electron-electron collision. In terms of collisions, the conditions in the solar interior are completely different from those of a planet atmosphere.

\subsection{Escape velocity from one proton}

However, even in the presence of collisions, the electron velocity values are  nearly all higher than the escape velocity as it results from gravity, as defined in Eq. (\ref{eq--gravity}). We also need to consider  the attracting effect exerted by protons on electrons, when the proton velocity remains generally lower than the escape velocity and the protons submitted to the star gravity. We propose the following model to analyze  this effect.

We first assume local electric neutrality and the same number of mixed electrons and protons. Then we assume that their electric fields roughly balance two by two in an  electron-proton pair, even if they are not linked to each other. Thus, an escaping electron is submitted to the electric field of a single proton. The calculation of this electric field requires knowing the distance between the electron and the proton, which can be taken as the mean distance between particles, which is the 
% MathType!MTEF!2!1!+-
% faaahqart1ev3aaaKnaaaaWenf2ys9wBH5garuavP1wzZbItLDhis9
% wBH5garmWu51MyVXgaruWqVvNCPvMCaebbnrfifHhDYfgasaacH8sr
% ps0lbbf9q8WrFfeuY-ribbf9v8qqaqFr0xc9pk0xbba9q8WqFfea0-
% yr0RYxir-Jbba9q8aq0-yq-He9q8qqQ8frFve9Fve9Ff0dc9Gqpi0d
% meaabaqaciGacaGaaeqabaWaaeaaeaaakeaadaWcgaqaaiabgkHiTi
% aaigdaaeaacaaIZaaaaaaa!33FA!
${{ - 1} \mathord{\left/
 {\vphantom {{ - 1} 3}} \right.
 \kern-\nulldelimiterspace} 3}$
power of their density, which is roughly the same for electrons and protons. This electric field derives from a potential, as does the gravitation field. Thus, there is also an escape velocity, as there is for gravitation. The escape velocity for freeing the electron from the corresponding single proton is then
\begin{equation}
% MathType!MTEF!2!1!+-
% faaahqart1ev3aaaKnaaaaWenf2ys9wBH5garuavP1wzZbItLDhis9
% wBH5garmWu51MyVXgaruWqVvNCPvMCaebbnrfifHhDYfgasaacH8sr
% ps0lbbf9q8WrFfeuY-ribbf9v8qqaqFr0xc9pk0xbba9q8WqFfea0-
% yr0RYxir-Jbba9q8aq0-yq-He9q8qqQ8frFve9Fve9Ff0dc9Gqpi0d
% meaabaqaciGacaGaaeqabaWaaeaaeaaakeaacaWG2bWaaSbaaSqaai
% aabwgacaqGZbGaae4yaaqabaGccqGH9aqpdaGcaaqaamaalaaabaGa
% aGOmaiaadghadaahaaWcbeqaaiaaikdaaaaakeaacaaI0aGaeqiWda
% NaeqyTdu2aaSbaaSqaaiaaicdaaeqaaOGaamyBamaaBaaaleaacaWG
% Lbaabeaakiaad6gadaahaaWcbeqaaiabgkHiTmaalyaabaGaaGymaa
% qaaiaaiodaaaaaaaaaaeqaaaaa!43FE!
{v_{{\rm{esc}}}} = \sqrt {\frac{{2{q^2}}}{{4\pi {\varepsilon _0}{m_e}{n^{ - {1 \mathord{\left/
 {\vphantom {1 3}} \right.
 \kern-\nulldelimiterspace} 3}}}}}} \ \ ,
\end{equation}
where $q$ is the elementary charge, $m_e$ is the electron mass, and $n$ the density, which is nearly the same for electrons and protons.

In the solar interior conditions given by \citet{Allen-73}, we then obtain that this escape velocity is nevertheless six times lower than the quadratic average thermal velocity for electrons, which then also escape from the protons, at least initially. As the escape velocity is obtained by equating the particle kinetic and potential energies, the total escape velocity from gravity and proton  for an electron is
\begin{equation}
% MathType!MTEF!2!1!+-
% faaahqart1ev3aaaKnaaaaWenf2ys9wBH5garuavP1wzZbItLDhis9
% wBH5garmWu51MyVXgaruWqVvNCPvMCaebbnrfifHhDYfgasaacH8sr
% ps0lbbf9q8WrFfeuY-ribbf9v8qqaqFr0xc9pk0xbba9q8WqFfea0-
% yr0RYxir-Jbba9q8aq0-yq-He9q8qqQ8frFve9Fve9Ff0dc9Gqpi0d
% meaabaqaciGacaGaaeqabaWaaeaaeaaakeaacaWG2bWaaSbaaSqaai
% aabwgacaqGZbGaae4yaaqabaGccqGH9aqpdaGcaaqaamaalaaabaGa
% aGOmaiaadEeacaWGnbWaaSbaaSqaaiabgYda8aqabaaakeaacaWGsb
% aaaiabgUcaRmaalaaabaGaaGOmaiaadghadaahaaWcbeqaaiaaikda
% aaaakeaacaaI0aGaeqiWdaNaeqyTdu2aaSbaaSqaaiaaicdaaeqaaO
% GaamyBamaaBaaaleaacaWGLbaabeaakiaad6gadaahaaWcbeqaaiab
% gkHiTmaalyaabaGaaGymaaqaaiaaiodaaaaaaaaaaeqaaaaa!495B!
{v_{{\rm{esc}}}} = \sqrt {\frac{{2G{M_ < }}}{R} + \frac{{2{q^2}}}{{4\pi {\varepsilon _0}{m_e}{n^{ - {1 \mathord{\left/
 {\vphantom {1 3}} \right.
 \kern-\nulldelimiterspace} 3}}}}}}
,\end{equation}
and we obtain that for electrons it is 5.5 times lower than the quadratic average thermal velocity at 0.5 $R_{\sun}$. The electrons then initially escape from gravity and protons.

\subsection{Escape velocity from several protons}

As the electron cloud is  free at the beginning, in a first approximation it begins to expand. The result is a decrease in the electron density, while the proton density remains unchanged. As a consequence, the number of electrons and protons is not yet the same in a given volume. As a result, an escaping electron is now submitted to the attraction of more than one mean proton, depending on the density ratio. If we denote as $\rho$ this density ratio
\begin{equation}
% MathType!MTEF!2!1!+-
% faaahqart1ev3aaaKnaaaaWenf2ys9wBH5garuavP1wzZbItLDhis9
% wBH5garmWu51MyVXgaruWqVvNCPvMCaebbnrfifHhDYfgasaacH8sr
% ps0lbbf9q8WrFfeuY-ribbf9v8qqaqFr0xc9pk0xbba9q8WqFfea0-
% yr0RYxir-Jbba9q8aq0-yq-He9q8qqQ8frFve9Fve9Ff0dc9Gqpi0d
% meaabaqaciGacaGaaeqabaWaaeaaeaaakeaacqaHbpGCcqGH9aqpda
% Wcaaqaaiaad6gadaWgaaWcbaGaamiCaaqabaaakeaacaWGUbWaaSba
% aSqaaiaadwgaaeqaaaaaaaa!387C!
\rho  = \frac{{{n_p}}}{{{n_e}}} \ \ ,
\end{equation}
where $n_e$ and $n_p$ are the electron and proton density, respectively, the total escape velocity of the electron submitted to $\rho  > 1$ protons and to gravity becomes
\begin{equation}
% MathType!MTEF!2!1!+-
% faaahqart1ev3aaaKnaaaaWenf2ys9wBH5garuavP1wzZbItLDhis9
% wBH5garmWu51MyVXgaruWqVvNCPvMCaebbnrfifHhDYfgasaacH8sr
% ps0lbbf9q8WrFfeuY-ribbf9v8qqaqFr0xc9pk0xbba9q8WqFfea0-
% yr0RYxir-Jbba9q8aq0-yq-He9q8qqQ8frFve9Fve9Ff0dc9Gqpi0d
% meaabaqaciGacaGaaeqabaWaaeaaeaaakeaacaWG2bWaaSbaaSqaai
% aabwgacaqGZbGaae4yaaqabaGccqGH9aqpdaGcaaqaamaalaaabaGa
% aGOmaiaadEeacaWGnbWaaSbaaSqaaiabgYda8aqabaaakeaacaWGsb
% aaaiabgUcaRmaalaaabaGaaGOmaiabeg8aYnaaCaaaleqabaWaaSGb
% aeaacaaIYaaabaGaaG4maaaaaaGccaWGXbWaaWbaaSqabeaacaaIYa
% aaaaGcbaGaaGinaiabec8aWjabew7aLnaaBaaaleaacaaIWaaabeaa
% kiaad2gadaWgaaWcbaGaamyzaaqabaGccaWGUbWaa0baaSqaaiaadc
% haaeaadaWcgaqaaiabgkHiTiaaigdaaeaacaaIZaaaaaaaaaaabeaa
% aaa!4DD6!
{v_{{\rm{esc}}}} = \sqrt {\frac{{2G{M_ < }}}{R} + \frac{{2{\rho ^{{2 \mathord{\left/
 {\vphantom {2 3}} \right.
 \kern-\nulldelimiterspace} 3}}}{q^2}}}{{4\pi {\varepsilon _0}{m_e}n_p^{{{ - 1} \mathord{\left/
 {\vphantom {{ - 1} 3}} \right.
 \kern-\nulldelimiterspace} 3}}}}} \ \ ,
\end{equation}
because the mean distance between the electron and the other particles is
% MathType!MTEF!2!1!+-
% faaahqart1ev3aaaKnaaaaWenf2ys9wBH5garuavP1wzZbItLDhis9
% wBH5garmWu51MyVXgaruWqVvNCPvMCaebbnrfifHhDYfgasaacH8sr
% ps0lbbf9q8WrFfeuY-ribbf9v8qqaqFr0xc9pk0xbba9q8WqFfea0-
% yr0RYxir-Jbba9q8aq0-yq-He9q8qqQ8frFve9Fve9Ff0dc9Gqpi0d
% meaabaqaciGacaGaaeqabaWaaeaaeaaakeaacaWGUbWaa0baaSqaai
% aadwgaaeaadaWcgaqaaiabgkHiTiaaigdaaeaacaaIZaaaaaaaaaa!3604!
$n_e^{{{ - 1} \mathord{\left/
 {\vphantom {{ - 1} 3}} \right.
 \kern-\nulldelimiterspace} 3}}$.

If we denote as $x$ the ratio between the electron escape and thermal velocities
\begin{equation}
 % MathType!MTEF!2!1!+-
% faaahqart1ev3aaaKnaaaaWenf2ys9wBH5garuavP1wzZbItLDhis9
% wBH5garmWu51MyVXgaruWqVvNCPvMCaebbnrfifHhDYfgasaacH8sr
% ps0lbbf9q8WrFfeuY-ribbf9v8qqaqFr0xc9pk0xbba9q8WqFfea0-
% yr0RYxir-Jbba9q8aq0-yq-He9q8qqQ8frFve9Fve9Ff0dc9Gqpi0d
% meaabaqaciGacaGaaeqabaWaaeaaeaaakeaadaWcaaqaaiaadAhada
% WgaaWcbaGaaeyzaiaabohacaqGJbaabeaaaOqaaiaadAhadaWgaaWc
% baGaaeiDaiaabIgaaeqaaaaakiabg2da9iaadIhaaaa!3A9A!
\frac{{{v_{{\rm{esc}}}}}}{{{v_{{\rm{th}}}}}} = x \ \ ,
\end{equation}
we obtain from the preceding equations
\begin{equation}
% MathType!MTEF!2!1!+-
% faaahqart1ev3aaaKnaaaaWenf2ys9wBH5garuavP1wzZbItLDhis9
% wBH5garmWu51MyVXgaruWqVvNCPvMCaebbnrfifHhDYfgasaacH8sr
% ps0lbbf9q8WrFfeuY-ribbf9v8qqaqFr0xc9pk0xbba9q8WqFfea0-
% yr0RYxir-Jbba9q8aq0-yq-He9q8qqQ8frFve9Fve9Ff0dc9Gqpi0d
% meaabaqaciGacaGaaeqabaWaaeaaeaaakeaacqaHbpGCcqGH9aqpda
% WcaaqaaiaadIhadaahaaWcbeqaaiaaiodaaaaakeaacaWGUbWaa0ba
% aSqaaiaadchaaeaadaWcgaqaaiaaigdaaeaacaaIYaaaaaaaaaGcda
% qadaqaamaalaaabaGaaGinaiabec8aWjabew7aLnaaBaaaleaacaaI
% WaaabeaaaOqaaiaadghadaahaaWcbeqaaiaaikdaaaaaaOWaaSaaae
% aacaaIZaaabaGaaGOmaaaacaWGRbWaaSbaaSqaaiaabkeaaeqaaOGa
% amivaaGaayjkaiaawMcaamaaCaaaleqabaWaaSGbaeaacaaIZaaaba
% GaaGOmaaaaaaaaaa!488F!
\rho  = \frac{{{x^3}}}{{n_p^{{1 \mathord{\left/
 {\vphantom {1 2}} \right.
 \kern-\nulldelimiterspace} 2}}}}{\left( {\frac{{4\pi {\varepsilon _0}}}{{{q^2}}}\frac{3}{2}{k_{\rm{B}}}T} \right)^{{3 \mathord{\left/
 {\vphantom {3 2}} \right.
 \kern-\nulldelimiterspace} 2}}} \ \ ,
\end{equation}
which permits us to compute the proton-to-electron density ratio $\rho$ from the desired velocity ratio $x$. The electron escaping will certainly be suppressed when $x = 3$, which  means that the escape velocity will be at least three standard deviations higher than the thermal velocity for each velocity component. In this case, $\rho$ is found to be $6 \times 10^3$ at 0.5 $R_{\sun}$ in the \citet{Allen-73} conditions. This corresponds to a mean distance between electrons 18 times larger than that between protons. These quantities are found to be quite constant along the solar radius.

\subsection{Spreading time}

One important question is where exactly   do the initially free electrons escape, and the answer is to the distance where the electron cloud density is divided by the ratio $\rho$ defined above. We propose that the density scale height is the typical dimension of the cloud. We denote as $h$ this density scale height. Then the cloud extends at most up to $18/2 = 9$ times $h$ from its initial position. The division by two comes from the fact that the electron cloud may extend in two directions  from its initial position. The density scale height $h$ is on the order of 55,000 km inside the Sun, but sharply decreases at the surface, up to 110 km (pressure scale height) in the photosphere.

The time for the initially free electron cloud to reach these stopping distances can be evaluated from the Fick's laws of diffusion. If we denote as $D$ the diffusion coefficient, which is
\begin{equation}
% MathType!MTEF!2!1!+-
% faaahqart1ev3aaaKnaaaaWenf2ys9wBH5garuavP1wzZbItLDhis9
% wBH5garmWu51MyVXgaruWqVvNCPvMCaebbnrfifHhDYfgasaacH8sr
% ps0lbbf9q8WrFfeuY-ribbf9v8qqaqFr0xc9pk0xbba9q8WqFfea0-
% yr0RYxir-Jbba9q8aq0-yq-He9q8qqQ8frFve9Fve9Ff0dc9Gqpi0d
% meaabaqaciGacaGaaeqabaWaaeaaeaaakeaacaWGebGaeyypa0JaeS
% 4eHWMaamODamaaBaaaleaacaqG0bGaaeiAaaqabaaaaa!3788!
D = \ell {v_{{\rm{th}}}} \ \ ,
\end{equation}
where the quadratic average particle velocity is the thermal velocity $v_{\rm{th}}$ and $\ell$ is the shortest electron collision mean free path discussed above, the time $t$ necessary to reach the distance $d$ is
\begin{equation}
% MathType!MTEF!2!1!+-
% faaahqart1ev3aaaKnaaaaWenf2ys9wBH5garuavP1wzZbItLDhis9
% wBH5garmWu51MyVXgaruWqVvNCPvMCaebbnrfifHhDYfgasaacH8sr
% ps0lbbf9q8WrFfeuY-ribbf9v8qqaqFr0xc9pk0xbba9q8WqFfea0-
% yr0RYxir-Jbba9q8aq0-yq-He9q8qqQ8frFve9Fve9Ff0dc9Gqpi0d
% meaabaqaciGacaGaaeqabaWaaeaaeaaakeaacaWGKbGaeyypa0ZaaO
% aaaeaacaaI2aGaamiraiaadshaaSqabaaaaa!360B!
d = \sqrt {6Dt} \ \ .
\end{equation}
The diffusion velocity $v_{\rm{diff}}$ can then be defined as 
% MathType!MTEF!2!1!+-
% faaahqart1ev3aaaKnaaaaWenf2ys9wBH5garuavP1wzZbItLDhis9
% wBH5garmWu51MyVXgaruWqVvNCPvMCaebbnrfifHhDYfgasaacH8sr
% ps0lbbf9q8WrFfeuY-ribbf9v8qqaqFr0xc9pk0xbba9q8WqFfea0-
% yr0RYxir-Jbba9q8aq0-yq-He9q8qqQ8frFve9Fve9Ff0dc9Gqpi0d
% meaabaqaciGacaGaaeqabaWaaeaaeaaakeaacaWG2bWaaSbaaSqaai
% aabsgacaqGPbGaaeOzaiaabAgaaeqaaOGaeyypa0ZaaSGbaeaacaWG
% KbaabaGaamiDaaaaaaa!3953!
${v_{{\rm{diff}}}} = {d \mathord{\left/
 {\vphantom {d t}} \right.
 \kern-\nulldelimiterspace} t}$
and it is found that this diffusion velocity is very weak. The cloud slowly spreads out. The time needed to reach the $9h$ distance is found to be greater than the age of the  Universe for electrons of the more internal layers where $h$ is about 55,000 km. Thus, these electrons do not leave the star. Only the electrons lying in the layers close to the surface can reach their limit distance, and this distance $9h$ is also found  close to the surface because the density scale height is much smaller there. As a result, the electrons  do not leave the star, and there is an accumulation of them in the surface layer. The surface electrons reach their limit distance $9h$ at a velocity of $1.3 \times 10^{-6}$ m/s, which is 1.7 m/year. Thus, the spreading is quasi-static.

\subsection{Surface electron density}

We finally propose an attempt to evaluate the surface electron density. If we consider that the density scale height is roughly the length along which the density is divided by three, and if we also neglect for a moment the height variation of the density scale height $h$ and assume a constant $h$, the number of electrons per volume unit will be obtained by summing the number of electrons in each of the nine  internal layers, each expanded up to $9h$, and each being three times more populated than its more external neighbor. The total number of electrons in the surface layer is then increased by a factor
\begin{equation}
% MathType!MTEF!2!1!+-
% faaahqart1ev3aaaKnaaaaWenf2ys9wBH5garuavP1wzZbItLDhis9
% wBH5garmWu51MyVXgaruWqVvNCPvMCaebbnrfifHhDYfgasaacH8sr
% ps0lbbf9q8WrFfeuY-ribbf9v8qqaqFr0xc9pk0xbba9q8WqFfea0-
% yr0RYxir-Jbba9q8aq0-yq-He9q8qqQ8frFve9Fve9Ff0dc9Gqpi0d
% meaabaqaciGacaGaaeqabaWaaeaaeaaakeaacaaIXaGaey4kaSIaaG
% 4maiabgUcaRiaaiodadaahaaWcbeqaaiaaikdaaaGccqGHRaWkcqWI
% VlctcaaIZaWaaWbaaSqabeaacaaI4aaaaOGaeyypa0ZaaSaaaeaaca
% aIZaWaaWbaaSqabeaacaaI5aaaaOGaeyOeI0IaaGymaaqaaiaaioda
% cqGHsislcaaIXaaaaiabgIKi7kaaigdacaaIWaWaaWbaaSqabeaaca
% aI0aaaaaaa!45DC!
1 + 3 + {3^2} +  \cdots {3^8} = \frac{{{3^9} - 1}}{{3 - 1}} \approx {10^4} \ \ .
\end{equation}
However, on the surface, a part of these electrons associate with protons to form neutral hydrogen atoms. The initial proton density was $\rho$ times the electron density of the first expanded layer, with $\rho$ about $6 \times 10^3$ at 0.5 $R_{\sun}$, but similarly $5 \times 10^3$ in the more external layer in the \citet{Allen-73} conditions. Thus, the remaining free electron density is  comparable to the neutral hydrogen density. In Sect. \ref{sect--fields}, we  show that this is the order of magnitude where the resulting magnetization divergence can explain the observed magnetic field divergence.

\section{Electron density measurements in the solar photosphere}
\label{sect--measurements}

\begin{figure}
\resizebox{\hsize}{!}{\includegraphics{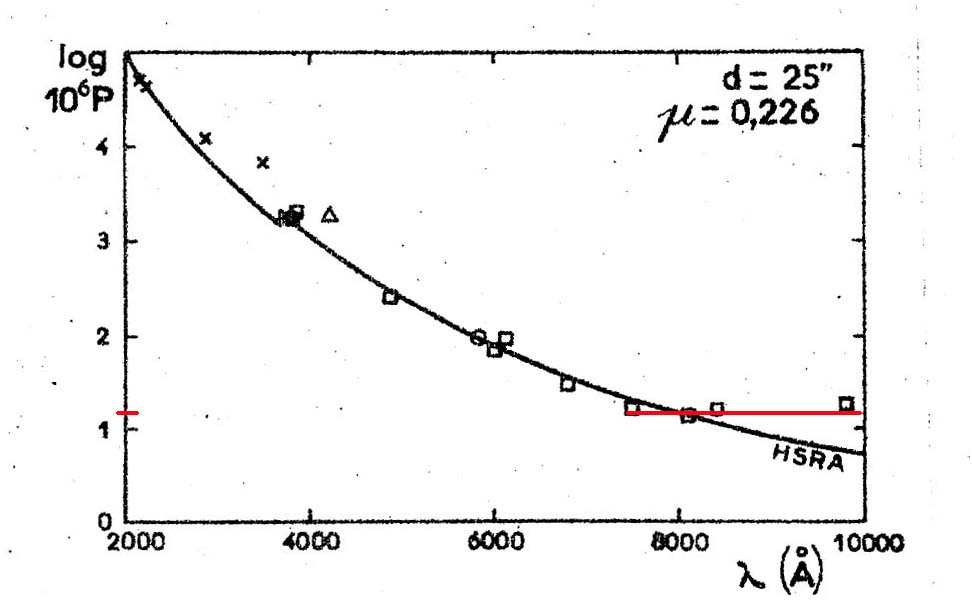}}
\caption{Linear polarization degree with polarization direction parallel to the solar limb, observed 25 arcsec inside the limb in the continuum and far from active regions, as a function of the wavelength. Observations: squares $\Box$ \citet{Leroy-72,Leroy-74}, triangles $\triangle$ \citet{Wiehr-75}, circles $\circ$ \citet{Mickey-Orrall-74}, crosses $\times$ \citet{Goutail-78}. Line: Model result based on the HSRA atmosphere model \citep{Dumont-Pecker-71}. From \citet{Leroy-77}, used with permission. Red line: starting from 8000 \AA, the polarization degree ceases to decrease following Rayleigh scattering on neutral hydrogen. It remains constant, as in Thomson scattering on electrons;   HSRA modeling would give Thomson scattering dominant only after 10,000 \AA.}
\label{Leroy}
\end{figure}

To our knowledge, the electron density measurements in the photosphere are only indirect. Results are derived from   spectroscopic analysis of the observed spectra  \citep[e.g., the models by][]{Vernazza-etal-81}. The hypothesis and equation of electric neutrality are   the basis of the analysis. The electron density then results, within this hypothesis, from the ionization equilibria of the different species present in the atmosphere. 

The most direct measurement we know is the investigation of the continuum linear polarization observed close to the solar limb and far from active regions, as done by Jean-Louis Leroy at the Pic-du-Midi Observatory in the 1970s \citep{Leroy-72,Leroy-74}. In \citet[Fig. 3, p. 167]{Leroy-77}, he presented the measurement and modeling synthesis, which we reproduce here in Fig. \ref{Leroy} with his permission. New measurements were made later on by \citet{Wiehr-78} (who improved the results of \citealt{Wiehr-75} included in Fig. \ref{Leroy}) and \citet{Wiehr-Bianda-03} (in excellent agreement with \citealt{Wiehr-78}), but in such  good agreement with the measurements and with the theoretical model presented in Fig. \ref{Leroy} that we did not find it necessary to redo the figure. These more recent measurements fall in the visible range and not in the infrared range, which is the one covered in  our discussion.

The continuum limb polarization results from Rayleigh scattering by the neutral hydrogen atoms. Rayleigh scattering is wavelength-dependent, which explains the decrease in the polarization degree as a function of increasing wavelength. The line in the figure is the result of a model by \citet{Debarbat-etal-70} and improved by \citet{Dumont-Pecker-71}, based on the HSRA model atmosphere. More recently, for the purpose of stellar studies, \citet{Kostogryz-Berdyugina-15} redid the calculations but with a series of model atmospheres including those of \citet{Vernazza-etal-81} and the HSRA model. Figure 5 of \citet{Kostogryz-Berdyugina-15} shows that all models lead to very close results for the theoretical limb polarization.

However, the observed polarization ceases to decrease with increasing wavelength at about 8000 \AA. This suggests that Thomson scattering on free electrons then becomes dominant. Thomson scattering is wavelength independent. This is indicated by the horizontal red line in Fig. \ref{Leroy}. However, the theoretical results based on the HSRA model would obtain a dominant Thomson scattering only after 10,000 \AA. Thus, observations seem to reveal an electron overdensity, with respect to the HSRA model, which is based on electric neutrality. Even if the orders of magnitude do not seem to exactly correspond to our hypothesis of comparable free electron and neutral hydrogen atom densities, this is  a confirmation (to be investigated by new observations and modeling) of our hypothesis of increased free electron density that can explain the difference between vertical and horizontal magnetic field gradients in the photosphere in and around sunspots via the $\vec{B}=\mu_0\left(\vec{H}+\vec{M}\right)$ law.

\section{Conclusion}
\label{sect--concl}

We have proposed an explanation of  the non-zero divergence of the observed magnetic field \citep[see the observation review by][]{Balthasar-18} by investigating the law $\vec{B}=\mu_0\left(\vec{H}+\vec{M}\right)$ in magnetized media. In order to obtain a non-negligible magnetization $\vec{M}$ at the surface of the Sun where the measurements are performed, we have invoked electron thermal escape from the more internal layers, where the electron thermal velocity noticeably exceeds their escape velocity. We have paid attention to the retaining role of protons and show that it does not totally prevent electrons from escaping. The result is an electron accumulation in the surface layer.

Positive free charges   could contribute in an additive manner to the magnetization. The positive charges gyrate about the induction or field in the opposite sense with respect to the negative charges \citep[see Fig. 2.8 of][]{Meyer-Vernet-07}, but as the charges are also opposite themselves, the elementary corresponding loop currents are similar for the two charge types, leading to their additive contribution to the magnetization. However, the usual photosphere models like VALC and HRSA are built on the hypothesis of electric neutrality. Then, if there is electric neutrality in the layer, the charge densities and then $\vec{M}$ remain low, following the spectroscopic analysis results. We cannot keep both non-negligible $\vec{M}$ and electric neutrality in the layer.

The escaping electrons would accumulate at the Sun's surface, when the protons would remain lower. Electric fields would then appear inside the star, but the electric effects would remain inside. As the global charge would remain zero or very weak, no electric effects would follow outside the star, in a first approximation.

The final result is that what is measured by Zeeman effect analysis is the magnetic field $\vec{H}$ and not the magnetic induction $\vec{B}$. However, the magnetohydrodynamical modeling requires the knowledge of $\vec{B}$. It can be deduced from the measurements $\vec{H}$ by applying the law $\vec{B}=\mu_0\left(\vec{H}+\vec{M}\right)$, provided $\vec{M}$ is also known. $\vec{M}$ would have to be reconstructed from its divergence $\rm{div}\vec{M}$, which is opposite to that of the magnetic field $\vec{H}$, because $\rm{div}\vec{B} = 0$, which implies $\rm{div}\vec{M} = - \rm{div}\vec{H}$, which is measured. The conduction currents $\vec{J}$,  however, can be directly derived from the measurements $\vec{H}$ because $\rm{curl}\vec{H} = \vec{J}$, as detailed in Eqs. (12-15).

\begin{acknowledgements}
I am deeply indebted to Jacques Dubau, Pascal D\'emoulin, Wan-\"U Lydia Tchang-Brillet, Nicole Meyer-Vernet for very fruitful discussions, for their suggestions, comments, analyses and provided references, also Silvano Bonazzola, Nicole Feautrier and Sylvie Sahal-Br\'echot. I am particularly indebted to Jean-Louis Leroy who pointed me the results of his observations cited in Sect. \ref{sect--measurements}, able to confirm the present analysis, and for fruitful discussions also. I am also deeply indebted to Pr Damien Gignoux, from Institut N\'eel, Emeritus Professor at Universit\'e Joseph Fourier, Grenoble (France), for introduction to magnetized material physics, helpful suggestions and a critical reading of the manuscript. G\'erard Belmont is also greatly thanked for previous discussions. I am grateful and indebted to an anonymous referee for helpful questions. This work was initiated by series of observations at the TH\'EMIS telescope of the French Centre National de la Recherche Scientifique (CNRS), thanks to its Programme National Soleil-Terre (PNST).
\end{acknowledgements}

\bibliographystyle{aa}
\bibliography{bommierrefs}

\begin{thebibliography}{35}
\expandafter\ifx\csname natexlab\endcsname\relax\def\natexlab#1{#1}\fi

\bibitem[{{Allen}(1973)}]{Allen-73}
{Allen}, C. 1973, Astrophysical Quantities, third edition edn. (University of
  London: The Athlone Press)

\bibitem[{{Balthasar}(2018)}]{Balthasar-18}
{Balthasar}, H. 2018, \solphys, 293, 120

\bibitem[{{Beck} \& {Meyer-Vernet}(2008)}]{Beck-MeyerV-08}
{Beck}, A. \& {Meyer-Vernet}, N. 2008, American Journal of Physics, 76, 934

\bibitem[{{Bommier}(2013)}]{Bommier-13}
{Bommier}, V. 2013, Physics Research International, Volume 2013 (2013), Article
  id.195403, 16 pages, 2013, 195403

\bibitem[{{Bommier}(2014)}]{Bommier-14}
{Bommier}, V. 2014, Comptes Rendus Physique, 15, 430

\bibitem[{{Bommier}(2015)}]{Bommier-15}
{Bommier}, V. 2015, in IAU Symposium, Vol. 305, Polarimetry, ed. K.~N.
  {Nagendra}, S.~{Bagnulo}, R.~{Centeno}, \& M.~{Jes{\'u}s Mart{\'{\i}}nez
  Gonz{\'a}lez}, 28--34

\bibitem[{{Bommier} {et~al.}(2007){Bommier}, {Landi Degl'Innocenti},
  {Landolfi}, \& {Molodij}}]{Bommier-etal-07}
{Bommier}, V., {Landi Degl'Innocenti}, E., {Landolfi}, M., \& {Molodij}, G.
  2007, \aap, 464, 323

\bibitem[{{Bruls} {et~al.}(1991){Bruls}, {Lites}, \& {Murphy}}]{Bruls-etal-91}
{Bruls}, J.~H.~M.~J., {Lites}, B.~W., \& {Murphy}, G.~A. 1991, in Solar
  Polarimetry, Proceedings of the 11st National Solar Observatory/Sacramento
  Peak Summer Workshop, Sunspot, New Mexico, 27-31 August 1990, ed. L.~J.
  {November} (National Solar Observatory), 444--456

\bibitem[{{Chamberlain}(1963)}]{Chamberlain-63}
{Chamberlain}, J.~W. 1963, \planss, 11, 901

\bibitem[{{Debarbat} {et~al.}(1970){Debarbat}, {Dumont}, \&
  {Pecker}}]{Debarbat-etal-70}
{Debarbat}, S., {Dumont}, S., \& {Pecker}, J.~C. 1970, \aap, 8, 231

\bibitem[{{Delcroix} \& {Bers}(1994)}]{Delcroix-Bers-94}
{Delcroix}, J.~L. \& {Bers}, A. 1994, Physique des Plasmas, Savoirs Actuels
  (Paris: Inter\'Editions / CNRS \'Editions)

\bibitem[{{du Tr\'emolet de Lacheisserie} {et~al.}(2002){du Tr\'emolet de
  Lacheisserie}, {Gignoux}, \& {Schlenker}}]{Gignoux-etal-02}
{du Tr\'emolet de Lacheisserie}, E., {Gignoux}, D., \& {Schlenker}, M. 2002,
  Magnetism I-Fundamentals (Dordrecht: Kluwer Academic Publishers)

\bibitem[{{Dumont} \& {Pecker}(1971)}]{Dumont-Pecker-71}
{Dumont}, S. \& {Pecker}, J.-C. 1971, \aap, 10, 118

\bibitem[{{Faurobert} {et~al.}(2009){Faurobert}, {Aime}, {P{\'e}rini},
  {Uitenbroek}, {Grec}, {Arnaud}, \& {Ricort}}]{Faurobert-etal-09}
{Faurobert}, M., {Aime}, C., {P{\'e}rini}, C., {et~al.} 2009, \aap, 507, L29

\bibitem[{{Garnier} {et~al.}(1998){Garnier}, {Gignoux}, {Schmitt}, \&
  {Shigeoka}}]{Garnier-etal-98}
{Garnier}, A., {Gignoux}, D., {Schmitt}, D., \& {Shigeoka}, T. 1998, \prb, 57,
  5235

\bibitem[{{Gignoux} \& {Schmitt}(1993)}]{Gignoux-Schmitt-93}
{Gignoux}, D. \& {Schmitt}, D. 1993, \prb, 48, 12682

\bibitem[{{Goutail}(1978)}]{Goutail-78}
{Goutail}, F. 1978, \aap, 64, 73

\bibitem[{{Jackson}(1975)}]{Jackson-75}
{Jackson}, J.~D. 1975, Classical Electrodynamics, second edition edn. (New
  York: John Wiley \& Sons)

\bibitem[{{Khomenko} \& {Collados}(2007)}]{Khomenko-Collados-07}
{Khomenko}, E. \& {Collados}, M. 2007, \apj, 659, 1726

\bibitem[{{Kostogryz} \& {Berdyugina}(2015)}]{Kostogryz-Berdyugina-15}
{Kostogryz}, N.~M. \& {Berdyugina}, S.~V. 2015, \aap, 575, A89

\bibitem[{{Landau} \& {Lifshitz}(1973)}]{Landau-Lifshitz-84}
{Landau}, L.~D. \& {Lifshitz}, E.~M. 1973, Electrodynamics of continuous media,
  second edition edn. (Oxford: Pergamon Press)

\bibitem[{{Langevin}(1905)}]{Langevin-05}
{Langevin}, P. 1905, Annales de chimie et de physique, 5, 70

\bibitem[{{Leroy}(1972)}]{Leroy-72}
{Leroy}, J.~L. 1972, \aap, 19, 287

\bibitem[{{Leroy}(1974)}]{Leroy-74}
{Leroy}, J.~L. 1974, \solphys, 36, 81

\bibitem[{{Leroy}(1977)}]{Leroy-77}
{Leroy}, J.~L. 1977, in Reports from the Observatory of Lund, Vol.~12,
  Measurements and interpretation of polarization arising in the solar
  chromosphere and corona : proceedings of a workshop held at Lund Observatory,
  May 9-13, 1977, ed. J.~O. {Stenflo} (Lund Observatory), 161--170

\bibitem[{{Maltby} {et~al.}(1986){Maltby}, {Avrett}, {Carlsson},
  {Kjeldseth-Moe}, {Kurucz}, \& {Loeser}}]{Maltby-etal-86}
{Maltby}, P., {Avrett}, E.~H., {Carlsson}, M., {et~al.} 1986, \apj, 306, 284

\bibitem[{{Meyer-Vernet}(2007)}]{Meyer-Vernet-07}
{Meyer-Vernet}, N. 2007, Basics of the Solar Wind (Cambridge University Press)

\bibitem[{{Mickey} \& {Orrall}(1974)}]{Mickey-Orrall-74}
{Mickey}, D.~L. \& {Orrall}, F.~Q. 1974, \aap, 31, 179

\bibitem[{{Solanki}(2003)}]{Solanki-03}
{Solanki}, S.~K. 2003, \aapr, 11, 153

\bibitem[{{Vernazza} {et~al.}(1981){Vernazza}, {Avrett}, \&
  {Loeser}}]{Vernazza-etal-81}
{Vernazza}, J.~E., {Avrett}, E.~H., \& {Loeser}, R. 1981, \apjs, 45, 635

\bibitem[{{Weiss}(1907)}]{Weiss-07}
{Weiss}, P. 1907, Journal de Physique Th\'eorique et Appliqu\'ee, 6, 661

\bibitem[{{Wiehr}(1975)}]{Wiehr-75}
{Wiehr}, E. 1975, \aap, 38, 303

\bibitem[{{Wiehr}(1978)}]{Wiehr-78}
{Wiehr}, E. 1978, \aap, 67, 257

\bibitem[{{Wiehr} \& {Bianda}(2003)}]{Wiehr-Bianda-03}
{Wiehr}, E. \& {Bianda}, M. 2003, \aap, 398, 739

\bibitem[{{Zhang} {et~al.}(1994){Zhang}, {Gignoux}, {Schmitt}, {Franse},
  {Kayzel}, {Kim-Ngan}, \& {Radwanski}}]{Zhang-etal-94}
{Zhang}, F.~Y., {Gignoux}, D., {Schmitt}, D., {et~al.} 1994, Journal of
  Magnetism and Magnetic Materials, 130, 108

\end{thebibliography}

\end{document}